\documentclass[10pt, journal]{IEEEtran}

\usepackage{amsthm} %defined already in ieeeconf
\usepackage{amsmath}    %For theorems
\IEEEoverridecommandlockouts
\usepackage{empheq}
\usepackage{bm}
\usepackage{soul}
\usepackage{graphicx}
\usepackage{indentfirst}
\usepackage{epstopdf}
\usepackage{amssymb}
\usepackage{url}
\usepackage{enumitem} %defined already in ieeeconf
\usepackage{multirow}
\usepackage{hhline}
\usepackage{booktabs}
\usepackage{mathtools}
\usepackage{makecell}
\usepackage[linesnumbered,boxed,commentsnumbered,ruled,vlined,longend]{algorithm2e}
\usepackage{comment}

\makeatother
\DeclareMathAlphabet\mathbfcal{OMS}{cmsy}{b}{n}



\usepackage{stackengine}

\newcommand{\mat}[1]{\boldsymbol{#1}}

\newcommand{\bmat}[1]{\begin{bmatrix} #1 \end{bmatrix}}

\providecommand{\mA}{\ensuremath{\mat{A}}}
\providecommand{\mB}{\ensuremath{\mat{B}}}
\providecommand{\mC}{\ensuremath{\mat{C}}}

\providecommand{\mE}{\ensuremath{\mat{E}}}

\providecommand{\mG}{\ensuremath{\mat{G}}}

\providecommand{\mI}{\ensuremath{\mat{I}}}

\providecommand{\mL}{\ensuremath{\mat{L}}}

\providecommand{\mO}{\ensuremath{\mat{O}}}
\providecommand{\mP}{\ensuremath{\mat{P}}}
\providecommand{\mQ}{\ensuremath{\mat{Q}}}

\providecommand{\mS}{\ensuremath{\mat{S}}}
\providecommand{\mT}{\ensuremath{\mat{T}}}

\providecommand{\mV}{\ensuremath{\mat{V}}}
\providecommand{\mW}{\ensuremath{\mat{W}}}
\providecommand{\mX}{\ensuremath{\mat{X}}}

\newcommand{\m}{\boldsymbol}
\allowdisplaybreaks[4]
\usepackage[colorlinks = true,
linkcolor = blue,
urlcolor  = blue,
citecolor = blue,
anchorcolor = blue]{hyperref}

\newcommand{\mc}[1]{\mathcal{#1}}
\newcommand{\mbb}[1]{\mathbb{#1}}
\newcommand{\mr}[1]{\mathrm{#1}}
\usepackage[framemethod=TikZ]{mdframed}
\mdfdefinestyle{MyFrame}{%
	linecolor=black,
	outerlinewidth=1.25pt,
	roundcorner=1.25pt,
	innerrightmargin=5pt,
	innerleftmargin=5pt,}
	
\usepackage[noabbrev]{cleveref}

\usepackage{mathtools}

\DeclarePairedDelimiter\abs{\lvert}{\rvert}%
\DeclarePairedDelimiter\norm{\lVert}{\rVert}%

\makeatletter
\let\oldabs\abs
\def\abs{\@ifstar{\oldabs}{\oldabs*}}
\let\oldnorm\norm
\def\norm{\@ifstar{\oldnorm}{\oldnorm*}}
\makeatother

\usepackage[english]{babel}
\usepackage[utf8]{inputenc}
\usepackage[super]{nth}

\usepackage{float}
\usepackage[caption = false]{subfig}

\usepackage{array}
\usepackage{threeparttable}

\usepackage[english]{babel}
\usepackage[utf8]{inputenc}
\usepackage[super]{nth}

\RequirePackage{filecontents}

\SetKwRepeat{Do}{do}{while}%

\usepackage{csquotes}

\usepackage{lipsum}
\usepackage{cite}
\usepackage{lettrine}

\usepackage{eqnarray,amsmath}
\usepackage{algpseudocode}

\usepackage[linesnumbered]{algorithm2e}% http://ctan.org/pkg/algorithm2e

\let\oldnl\nl% Store \nl in \oldnl
\newcommand{\nonl}{\renewcommand{\nl}{\let\nl\oldnl}}% Remove line number for one line

\title{\centering \huge {Structure-Preserving Model Order Reduction \\ for Nonlinear DAE Models of Power Networks}}

\author{Muhammad Nadeem and Ahmad F. Taha \vspace{-0.8cm}
	\thanks{The authors are with the Civil and Environmental Engineering Department, Vanderbilt University, 2201 West End Ave, Nashville, TN 37235, USA.
		Email addresses: muhammad.nadeem@vanderbilt.edu, ahmad.taha@vanderbilt.edu. This work is supported by National Science Foundation under Grants 2152450 and 2151571.}
}

\begin{document}

\newdimen\origiwspc%
\newdimen\origiwstr%
\origiwspc=\fontdimen2\font% original inter word space
\origiwstr=\fontdimen3\font% original inter word stretch

\fontdimen2\font=0.63ex% inter word space
%\fontdimen3\font=\origiwstr% inter word stretch

\maketitle

\markboth{IEEE Transactions on Power Systems, In Press, November 2024}{}

\begin{abstract}
This paper deals with the joint reduction of the number of dynamic and algebraic states of a nonlinear differential-algebraic equation (NDAE) model of a power network. The dynamic states depict the internal states of generators, loads, renewables, whereas the algebraic ones define network states such as voltages and phase angles. In the current literature of power system model order reduction (MOR), the algebraic constraints are usually neglected and the power network is commonly modeled via a set of ordinary differential equations (ODEs) instead of NDAEs. Thus, reduction is usually carried out for the dynamic states only and the algebraic variables are kept intact. This leaves a significant part of the system's size and complexity \textit{unreduced}. This paper addresses this aforementioned limitation by jointly reducing both dynamic and algebraic variables. As compared to the literature the proposed MOR techniques are endowed with the following features: \textit{(i)} no system linearization is required, \textit{(ii)} require no transformation to an equivalent or approximate ODE representation,  \textit{(iii)} guarantee that the reduced order model to be NDAE-structured and thus preserves the differential-algebraic structure of original power system model, and \textit{(iv)} can seamlessly reduce both dynamic and algebraic variables while maintaining high accuracy. Case studies performed on a 2000-bus power system reveal that the proposed MOR techniques are able to reduce system order while maintaining accuracy. 
 \end{abstract}

\vspace{-0.1cm}
\begin{IEEEkeywords}
Model order reduction, Balanced truncation, Nonlinear differential-algebraic equations models.
\end{IEEEkeywords}
\vspace{-0.4cm}
\section{Introduction and Paper Contributions}\label{section:intro}
\vspace{-0.1cm}
\lettrine[lines=2]{P}{o}wer systems form large-scale, complex networks that usually require large state-space expressions for accurate modeling. The complexity and size of power systems are even further increasing with the integration of renewables and other (power-electronics)-based distributed resources. Thus, the research area of model order reduction (MOR) in power systems is becoming highly crucial. Specifically, for the design of advanced feedback controllers (e.g., $\mathcal{H}_2$, $\mathcal{H}_\infty$, and LQR/LQG), the design of reduced-order model (ROM) is of extreme importance. This is because the order of these controllers matches the order of the system, thus for a very large system with thousands of state variables, the design of these controllers becomes intractable \cite{yogarathinam2017new, nadeem2023robust, nadeem2022dynamic}. 
%MOR techniques aim to significantly reduce the size of the model while preserving essential system behaviors such as controllability, observability, stability, and passivity. Reduced-order models (ROM)  provide a small-scale, simplified yet accurate representation of the original system which is invaluable in both research and practical applications.

The idea of model reduction is not new in power systems research, and significant relevant research has been proposed in the past two decades. Generally speaking, there are two main MOR philosophies in the current literature of power systems.  The first philosophy divides the power system into study (internal) and external areas. The study area is represented in detail, while the external area is simplified or approximated. This approach leverages coherency-based methods \cite{chow2013power, ma2011right, wang2005tracing, Germond}, which are rooted in the identification of coherent generator groups within the power system. The reduction process generally involves three steps: $(i)$ identifying coherency among generators, $(ii)$ dynamically reducing the system by aggregating the network and generators, and $(iii)$ potentially aggregating excitation controllers also in later stages. These coherency-based methods are highly regarded for their reliability in achieving dynamic equivalence in power systems. However, a notable limitation is the potential inability to reduce specific parts of the power network due to the inherent nature of coherency grouping \cite{qi2016nonlinear}.

The second philosophy draws from control theory literature, focusing on input-output-based model reduction methodologies.  These algorithms are theoretically robust and general purpose, making them suitable for a wide range of applications beyond traditional synchronous machines, including renewable resources. \textit{Our work focuses on such input-output-oriented MOR techniques}. These types of MOR methods are further classified into three main categories. The first category is based on Krylov subspace or moment-matching-based methods, which approximate the original system by matching moments of the system's transfer function. The second category relies on modal truncation-based methods like proper orthogonal decomposition (POD). These methods involve reducing the system by truncating less significant modes, based on their contribution to the system's dynamics. The third category forms balanced realizations or Gramian-based methods such as balanced-POD
(BPOD) and balanced truncation (BT). The aim herein is to reduce the system by identifying and retaining the most controllable and observable states---Gramians are matrices that quantify observability and controllability of dynamic systems. The readers are referred to \cite{Athanasios_MOR, brunton2019data} for further details about these techniques. 

These MOR methodologies have also been widely applied to power systems to construct various ROMs. For instance, authors in \cite{safaee2022structure, yogarathinam2017new} have proposed moment matching-based MOR techniques. In \cite{kaur2016challenges, freitas2008gramian, osipov2018adaptive}, researchers have proposed balanced realization-based MOR algorithms for linear ODE-based power system models. Later, these works have been extended in \cite{qi2016nonlinear} to propose MOR for nonlinear ODE-based power system models where instead of using Grammians, empirical controllability and observability covariance matrices are used to balance and truncate the system. Similarly, in \cite{rommes2006efficient, martins2007computation, martins2003computing} various modal truncation-based reduction algorithms have been proposed. Moreover, a Loewner matrix method-based approach for efficient model order reduction and system identification in power systems was recently proposed in \cite{rergis2018loewner}. Additionally, \cite{acle2019parameter} introduced a parametric MOR technique that preserves critical device parameters while effectively reducing large-scale power system models. Readers are referred to \cite{singh2011review, djukic2012dynamic} for a detailed survey of the existing MOR approaches in power systems.

%Moreover, recently in \cite{rergis2018loewner}  Loewner matrix method-based efficient model order reduction and system identification in power systems have been proposed. While authors in \cite{acle2019parameter} proposed a parametric MOR method that preserves essential device parameters while reducing large-scale power system models. 
However, in most of the current MOR power system literature, the algebraic constraints (modeling power/current balance) are usually neglected and the power system is modeled via a set of linear ODEs (or converted to linear ODEs) in order to apply MOR concepts from linear system control theory. The recent work in \cite{qi2016nonlinear} has considered nonlinearity in their design. Yet again, the algebraic constraints are neglected and thus MOR is only carried out for dynamic variables while keeping the algebraic variables intact. This is problematic as algebraic variables often constitute a large portion of the model, representing essential electrical quantities across the network. Ignoring these in the reduction process means a significant part of the system's complexity and size remains untouched, limiting the effectiveness of MOR. Also, considering the complete nonlinear differential-algebraic (NDAE) dynamics is essential because the linear/nonlinear ODE-based power system models cannot capture the effects of topological changes (like the tripping of transmission lines, etc.) \cite{GRO201612}. 

%Furthermore, it is unclear how to include dynamics of loads and renewables in the ODE-based models \cite{Wu2019InfluenceOL}.

The studies that focus on the DAE power system models such as \cite{freitas2008gramian, muntwiler2023stiffness} also require the conversion of power system models to an equivalent ODE representation. This is done by finding an explicit relationship of the algebraic variable (through the algebraic constraint model) and substituting it back into the dynamic system \cite{freitas2008gramian}. However, such equivalent ODE representation is only possible for linear-DAE (LDAE) systems and cannot be applied to NDAE power systems as there is no explicit equation for algebraic variables because of the presence of nonlinearity.  Notice that the power system algebraic variables are known to be highly nonlinear as the algebraic constraints are power/current balance equations which are characterized by trigonometric terms such as sines and cosines, reflecting the physical laws of electrical networks. Thus, algebraic states cannot be isolated (or expressed explicitly) and plugged back into the dynamic system as done in the case of the LDAE system in \cite{freitas2008gramian}.

Some studies have been carried out using Krylov-based methodology such as \cite{yogarathinam2017new} which does not require the conversion of system to an equivalent ODE and can directly be applied to the DAE system. However, they are also limited to LDAE power system models. Also, the MOR obtained using the Krylov-based method can be of higher order. This is because the constructed lower-order orthogonal basis depends on the number of system inputs, for example to match the first $l$ moments of the system,  the dimension of the reduced order orthogonal basis needs to be $l \times n_u$ (where $n_u$ are the number of inputs) \cite{rewienski2003trajectory}. Thus, in a system with a large number of inputs (such as power systems), the ROM can be of higher order.

\noindent \textbf{Paper Contributions.} \; In this paper, we present two MOR techniques that can directly be applied to the NDAE representation (without requiring power system models to be converted to equivalent ODE) of power systems and can reduce both dynamic and algebraic variables simultaneously to construct the corresponding ROM. The technical contributions are as follows:

%As both proposed techniques utilizes a completely different approach to design ROM, it is unclear how they would fare hence thorough numerical simulations are carried out to assess their performance. 

% designs the reduced order model based on the modes (or the energy content) in the time-domain transient simulation data while the second approach designs the ROM by balancing the empirical controllability and observability covariances. 
\begin{itemize}
	\item We propose two model reduction approaches for nonlinear power system models. One approach offers reducing the system order based on the modes (or the energy content) in the time-domain transient simulation data while the second approach, in contrast, offers designing ROM via balanced-realization (using empirical controllability and observability covariances). Since both proposed techniques adopt distinct approaches to designing ROM, it is unclear how they perform in terms of realizing full system dynamics.  As compared to \cite{yogarathinam2017new, freitas2008gramian}, the proposed methods do not rely on system linearization and/or equivalent ODE transformation---and as compared to \cite{qi2016nonlinear}, the proposed techniques can simultaneously reduce algebraic variables with dynamic variables. 
	\item Due to the diagonal structure of the proposed coordinate transformation matrix, the presented techniques ensure that the ROMs are NDAEs similar to the original power system model. This preserves the differential algebraic structure of power networks. This also	allows a seamless transition from reduced order to the full order dynamic and algebraic state variables.
    \item Thorough time-domain simulations under various transient conditions have been conducted to evaluate the performance of the proposed techniques. The test system includes: (1) a modified IEEE 39-bus system that models detailed dynamics of conventional power plants, a solar plant operating in grid-forming mode, and algebraic constraint models, and (2) a 2000-bus Texas network incorporating comprehensive $11^{th}$-order models of conventional power plants. Additionally, to demonstrate the advantages of the proposed MOR techniques, a comparison with the commonly used linear ODE-based balanced truncation method is presented.

	%\item Thorough time-domain simulation studies under different transient conditions have been carried out to assess the performance of the proposed techniques. The considered test system includes: \textit{(i)} modified IEEE 39-bus system modeling detailed conventional power plants dynamics, solar plant acting in grid-forming mode, composite load dynamics, and algebraic constraint models, and \textit{(ii)} a 2000-bus Texas network with comprehensive $11^{th}$-order conventional power plant models. Furthermore, to showcase the advantage of the proposed MOR techniques a comparison with the commonly used linear ODE-based balance truncation method has also been presented.
	
\end{itemize}

\noindent \textbf{Paper Organization.} \; Sec.~\ref{sec: System model} presents the advanced power system model considered in this study. Sec.~\ref{sec: prelim} delineates the problem formulation and scope. Sec.~\ref{sec:mor1} and~\ref{sec:mor2} present the two MOR algorithms for NDAE power system models. Case studies are presented in Sec.~\ref{sec:case studies} while the paper is concluded in Sec. \ref{sec: conclusion}. 

\vspace{-0.01cm}
\noindent {\textbf{Notation.}} \; Capital bold letters are used to represent matrices, as in $\m{A}$ while small capital bold letters, such as $\m{b}$, denote vectors. 
All the sets are represented in calligraphic fonts, such as $\mathcal{M}$ or $\mathcal{R}$. The symbol $\mathbb{R}^{u\times v}$ represents a real-valued matrix of size $u \times v$, similarly, $\mathbb{R}^{k}$ denotes a real-valued column vector with $k$ elements. The notations  $\m O$ and $\m I$ denote zero and identity matrix of appropriate dimensions, respectively. The union of two sets is denoted by $\cup$. The symbol $ \mathbb{S}^{n\times n}_{++}$  represents a square positive definite matrix of size $n \times n$ while the notation $\m e_j$ represents a column vectors of zeros with $1's$ only at location $j$. Also, all quantities are given in per unit (p.u) unless otherwise specified.
% For simplicity, some equations may omit the time dependency of vectors, indicating that $\m{x}_d(t)$, which would typically denote a vector dependent on time $t$, is simply written as $\m{x}_d$ in certain contexts. 
\vspace{-0.3cm}
\section{Nonlinear DAE Power System Model}\label{sec: System model}
\vspace{-0.1cm}
We consider a grid model with $S$ solar power plants, $G$ conventional power plants, and $L_k$ loads.  The overall power system is modeled as a graph with $\mathcal{N} = \left\lbrace 1,...,N\right\rbrace$ as the set of nodes/buses and $\mathcal{E}$ as the set of edges or transmission lines. The set of buses are grouped into various types: $\mathcal{R} = \left\lbrace 1,...,R\right\rbrace$ represents buses with PV power plants, $\mathcal{G} = \left\lbrace 1,...,G\right\rbrace$ denotes buses connected to the conventional power plants, $\mathcal{L}$ includes buses that contain loads, while  $\mathcal{U}$ collects non-unit buses that are not connected to any elements.

The overall grid model is mathematically represented using a set of NDAEs given as follows \cite{nadeem2023robust}:
\begin{subequations}~\label{equ:PSModel}
	%	\vspace{-0.2cm}
	\begin{align}
		\begin{split}
		\vspace{-0.5cm}
		\dot{\m x}(t) &= \m g(\m x_d(t), \;\m x_a(t),\;\m u(t), \;\m w(t)) ~\label{equ:PSModel-a} \end{split}\\
	\begin{split}
		\m 0 &= \m h(\m x_d(t),\;\m x_a(t), \;\m u(t),\;\m w(t)). ~\label{equ:PSModel-b}
			\end{split}
	\end{align}
	%\vspace{-0.1cm}
\end{subequations}
In model \eqref{equ:PSModel}, the set of differential equations \eqref{equ:PSModel-a} encompasses the dynamics of PV plants, conventional power plants, and composite loads dynamics, while the algebraic constraint \eqref{equ:PSModel-b} models the current/power balance equations of the electrical network. The vector  $\m x_d(t) \in \mbb{R}^{n_d}$ represents dynamic states and it lumps the dynamic variables of conventional power plants, PV plants, and composite load dynamics as $\m x_d = \bmat{\m x_G^\top& \m x_R^\top&\m x_L^\top}^\top$ where $\m x_G$ are the states of conventional power plants, $\m x_R$  denotes the states of PV plants, and $\m x_L$  represents the states of dynamic loads.  The notation $\m x_a(t) \in \mbb{R}^{n_a}$ denotes algebraic states and it contains the states of the network (voltage and current phasors). The vector $\m w(t) = \bmat{\m P_d^\top& \m I_s^\top}^\top\in \mbb{R}^{n_w}$ contains load demand $\mP_d$ and sun irradiance $\m I_s$ while $\m u(t)= \bmat{\m u_G^\top& \m u_R^\top}^\top \in \mbb{R}^{n_u}$ defines the system control inputs with $\m u_G$ and  $\m u_R$  denoting the control inputs of generators and solar farms, respectively. Further detailed explanations of these vectors and complete dynamical equations (set of differential equations)  describing the models of PV plants, conventional power plants, composite load dynamics, and system algebraic constraint model used in this study are given in Appendix \ref{appndix:ninth Gen_dynamics}.

By considering $\m x(t) = \bmat{\m x_d^\top & \m x_a^\top}^\top \in\mbb{R}^{n}$ as the overall state vector and $\m y(t)  \in\mbb{R}^{p}$ as the system output we can rewrite the electrical grid model \eqref{equ:PSModel} in the following compact format:
\begin{subequations}\label{eq:final_NDAE}
\begin{align}
	\m E\dot{{\m x}} &= {\m A}{\m x} +{{{\m  B}}_u} {{\m u} } + {\m f}\left({\m x},{\m u},{\m w} \right) + {\m B}_w \m w\\
	\m y &= \m C \m x
\end{align}
\end{subequations}
where  $\m E \in\mbb{R}^{n\times n}$ is a singular binary matrix encoding system algebraic model with rows of zeros, function ${\m f}\left({\m x},{\m u},{\m w} \right)$ represents the corresponding nonlinearity, $\mC \in\mbb{R}^{p\times n}$ is the output matrix,  while the rest of the real-valued matrices  ${\m B}_u \in\mbb{R}^{n\times n_u}$, ${\m A}\in\mbb{R}^{n\times n}$, ${\m B}_w \in\mbb{R}^{n\times n_w}$ maps the system control inputs $\m u$, state vector $\m x$,  and the disturbance vector $\m w$ in the power system dynamics. Furthermore, throughout the paper, we assume that in the NDAE model \eqref{eq:final_NDAE}, the pair $(\m E,\mA)$ is regular, and the power system model is observable and controllable. These assumptions are common and power system NDAEs are known to be regular, controllable, and observable \cite{nadeem2023robust, AranyaICSM2019}.
\vspace{-0.35cm}
\section{Preliminaries and Problem Description}\label{sec: prelim}
\vspace{-0.1cm}
Generally speaking, the model reduction process involves transforming the original high-dimensional system into a new coordinate system where the states are ordered based on their \textit{importance} (defined through balancing controllability and observability in the defined transformations or dominance in mode-based MOR). This transformation enables the identification and retention of the most significant states while discarding those with minimal impact on the system's input-output behavior. The outcome is a reduced-order model that approximates the behavior of the original system with far fewer states---making the ROMs more amenable to real-time control and state estimation. 

Having said that, to perform MOR let $\m x(t)= \mW \tilde{\m x}(t)$ be the coordinate transformation with $\mW  \in\mbb{R}^{n\times n}$ representing the non-singular transformation matrix and $\tilde{\m x} \in\mbb{R}^{n}$ as the new set of coordinates where states are hierarchically ordered. Then one can simply truncate $\m W$ as $\mW_R = \m W\m T  \in\mbb{R}^{n\times r}$ with  $\mT = \bmat{\mI&\mO}^\top \in\mbb{R}^{n\times r}$ and thus choose the first $r << n$ dominant states of the transformed system while removing the rest. This dramatically reduces the model's complexity while retaining most of the system input-output behavior. To construct the reduced order model, the Galerkin projection \cite{brunton2019data} is commonly used, which involves projecting the dynamics of the original system onto the subspace spanned by the retained states. For example, assuming appropriate $\mW_R$ for the NDAE model \eqref{eq:final_NDAE} has been determined then the corresponding  ROM using Galerkin projection can be constructed as:
\begin{subequations}\label{eq:ROM_initial}
	\begin{align}
 \m E_r\dot{{\m z}} &= \m A_r{\m z}  + \m B_{ur}{\m u} + \mW_L{\m f}\left({\m W_R \tilde{\m x}},{\m u},{\m w} \right) +\m B_{wr} \m w\\
		\m y &= \mC_r\m z
	\end{align}
\end{subequations}
where $\m z \in\mbb{R}^{r}$ represents the state of the reduced system and $\m W_L =  \mW_R^{-1} \in\mbb{R}^{r\times n}$ is the left side coordinate transformation matrix. The rest of the matrices in \eqref{eq:ROM_initial} are given as follows:
\begin{subequations}\label{eq:ROM_Mat}
\begin{align}
\m A_r &= \mW_L{\m A}\m W_R,\; \m B_{ur} = \mW_L{{\m  B}},\; \m B_{wr} = \mW_L{\m B}_w\\ \m C_r &= \m C \mW_R,\;\m E_r = \mW_L\m E\mW_R.
\end{align}
\end{subequations}
Throughout this paper, the subscript $r$ is used to represent the parameters associated with ROM. 
Consequently, the main objective of the paper is to design appropriate coordinate transformation $\mW$ and truncation matrix $\mT$ for the complete NDAE power system \eqref{eq:final_NDAE}, and then construct a structure-preserving (meaning $\mE_r$ needs to be singular and thus the ROM \eqref{eq:ROM_initial} should remain NDAE similar to the full order model) reduced model that retains the same input-output behavior, while having significantly fewer number of states or equations than the original power system model \eqref{eq:final_NDAE}. The proposed MOR techniques are proposed in the next sections.

\vspace{-0.3cm}
\section{Structure-Preserving POD (SP-POD) MOR}\label{sec:mor1}
\vspace{-0.1cm}
Here we introduce SP-POD-based methodology to construct ROM for the NDAE power system model. Generally speaking, the POD-based MOR commonly consists of three main steps. Firstly, the system is simulated under transient conditions, and data is collected. Then, POD is applied to this data set to extract the most significant modes or features to construct a coordinate transformation matrix $\mW$. These modes are orthogonal functions that represent the system's dynamics in descending order of energy or variance. Essentially, POD seeks to find a basis that captures the most significant patterns in the data. Then finally, a reduced model of the system is constructed using Galerkin projection as discussed in Sec. \ref{sec: prelim}. 

Having said that, to propose a POD-based MOR technique for the NDAE power system models we do the following. First, time domain simulation for $20$sec of model \eqref{eq:final_NDAE} is carried out under transient conditions (by adding a step disturbance in load demand as discussed later in Sec. \ref{sec:case studies}) and the dynamic and algebraic states data are collected and stored separately as follows:
\begin{subequations}\label{eq:X data}
	\begin{align}
		\mX_d &= \bmat{
			\mid & \mid & & \mid & \mid\\
			{\m x}_{d_0} & {\m x}_{d_1} & \cdots & {\m x}_{d_{t-1}} & {\m x}_{d_t}\\
			\mid & \mid & & \mid & \mid\\} \label{eq:X_d data}\\
		\mX_a &= \bmat{
			\mid & \mid & & \mid & \mid\\
			{\m x}_{a_0} & {\m x}_{a_1} & \cdots & {\m x}_{a_{t-1}} & {\m x}_{a_t}\\
			\mid & \mid & & \mid & \mid \label{eq:X_a data}\\}
	\end{align}
\end{subequations}
% \begin{align}\label{eq:X data}
% 	\mX_d = \bmat{{\m x}_{d_0},\;\;\;\cdots\;\;\;,{\m x}_{d_{t-1}}& {\m x}_{d_t}},\;\;\mX_a = \bmat{{\m x}_{a_0},\;\;\;\cdots\;\;\;,{\m x}_{a_{t-1}}& {\m x}_{a_t}}
% \end{align}
where $\mX_d \in\mbb{R}^{n_d\times t}$ encapsulates the dynamic system data, $\mX_a \in\mbb{R}^{n_a\times t}$ contain the data for the algebraic variables while  ${\m x}_{d_0}$, ${\m x}_{a_0}$, and so on represents the trajectories of dynamic and algebraic states, respectively, at time step $0$ to the final time step $t$. 

Next, we find the POD modes of this data by computing the singular value decomposition (SVD) of matrices $\mX_d$ and $\mX_a$ separately as follows. For the dynamic system data applying SVD we get $\mX_d = \m W_d\m\Sigma_d\m \Lambda_d$, where $\m\Sigma_d \in\mbb{R}^{n_d\times n_d}$ is a diagonal matrix that contains the Hankel singular values (HSVs) in descending order while matrices $\m\Lambda_d \in\mbb{R}^{t\times n_d}$, $\mW_d \in\mbb{R}^{n_d\times n_d}$ contains the right and left singular vectors, respectively. The columns of $\mW_d$ are ordered hierarchically from most dominant to least and are referred to as the POD modes as they capture the most energetic patterns of the data.
Similarly, for $\mX_a$ we get $\mX_a = \m W_a\m\Sigma_a\m \Lambda_a$ with matrices $\m W_a \in\mbb{R}^{n_a\times n_a}$, $\m\Lambda_a \in\mbb{R}^{t\times n_a}$ containing the left and right singular vectors while $\m\Sigma_a \in\mbb{R}^{n_a\times n_a}$ lumping the corresponding singular values. After determining the POD modes, we construct the final non-singular coordinate transformation matrix $\mW$ as follows:
\begin{align}\label{eq: W mat}
	\mW = \mr{blkdiag}(\mW_d,\;\mW_a)
\end{align} 
% \begin{align}\label{eq: W mat}
% 	\mW = \bmat{\mW_d&\m0\\\m0&\mW_a}.
% \end{align} 
\noindent where  $\mr{blkdiag}$ construct a block diagonal matrix. The designed $\mW$ can transform the  NDAE system \eqref{eq:final_NDAE} to a new set of coordinates where the states are ordered hierarchically from most important/dominant to least. It is worth mentioning that because of the block diagonal structure, the designed coordinated transformation matrix $\mW$ guarantees
that the transformed model is always an NDAE similar to the original system. Thus preserving the essential structure of the original system and hence one can easily move from one coordinate to another using $\m x= \mW \tilde{\m x}$ since $\mW$ is non-singular. Now, as mentioned earlier in the transformed coordinate the states are ordered, and then to construct the ROM we need to truncate the least important states. Hence, we construct the appropriate truncation matrix $\mT$ as follows:
\begin{align}\label{eq:T equation}
	\hspace{-0.1cm}\mT_d \hspace{-0.05cm}= \hspace{-0.05cm}\bmat{\mI_{dr}&\m0}^\top\hspace{-0.05cm}, \mT_a \hspace{-0.05cm}= \hspace{-0.05cm}\bmat{\mI_{ar} & \m0}^\top\hspace{-0.05cm},
	\mT \hspace{-0.05cm}=\hspace{-0.05cm}\mr{blkdiag}(\mT_d,\;\mT_a)
\end{align}
where  $\mT_d \in\mbb{R}^{n_d\times n_d}$, and $\mT_a \in\mbb{R}^{n_a\times n_a}$. In \eqref{eq:T equation} the dimension of $\mI_{dr}$ can be determined by examining the magnitude of HSVs contained in $\m\Sigma_d$, similarly $\m I_{ar}$ can be designed based on HSVs in $\m\Sigma_a$.
\vspace{-0.0cm}
\begin{algorithm}\label{alg:Algorithm 1}
	\caption{\text{SP-POD for power system NDAE models }}
	\DontPrintSemicolon 
     \textbf{Input:} NDAE~\eqref{eq:final_NDAE} parameters $\mA$, $\mB_u$, $\mB_{w}$, $\mE$, $\m f(\cdot)$, and $\m x_0$\;
    \textbf{Output:} ROM parameters $\mE_r$, $\mA_r$, $\mB_{r}$, $\mB_{wr}$, and $\m f_r(\cdot)$\;
    Create snapshot matrices $\mX_d$ \eqref{eq:X_d data}, $\mX_a$ \eqref{eq:X_a data}, and $\mX_f$ \eqref{eq:f data}\;
	\If{$n << t$}{Perform SVD as: \;
		$\mX_d = \m W_d\m\Sigma_d\m \Lambda_d$,\,\,
		$\mX_a = \m W_a\m\Sigma_a\m \Lambda_a$}
	\Else {Perform eigenvalue decomposition:\; 
		$\mX_d^\top\mX_d\mV_d= \m V_d\m \lambda_d$, \, \, $\mX_a^\top\mX_a\mV_a= \m V_a\m \lambda_a$\;
		Then design $\mW_d$ and $\mW_a$ as folows:\;
		$\mW_{d} =  \mX_d\mV_d\m\lambda_d^{\frac{-1}{2}}$, \,\, $\mW_{a} =  \mX_a\mV_a\m\lambda_a^{\frac{-1}{2}}$} 
	Construct $\mW = \mr{blkdiag}(\mW_d,\; \mW_a)$ as in \eqref{eq: W mat}\;
	Design $\mT_d$, $\mT_a$ \eqref{eq:T equation} by examining HSVs in  $\m\Sigma_d$ and $\m\Sigma_a$\;
	Construct $\mT = \mr{blkdiag}(\mT_d,\; \mT_a)$ as in \eqref{eq:T equation}\;
	Design $\mW_R = \mT\mW$ and  $\mW_L = \mW_R^{-1}$\;
	SVD $\mX_f$ \eqref{eq:f data} as:  $\m X_f = \mW_f\m\Sigma_f\m\Lambda_f$\;
	Start greedy algorithm to design $\mP_M$\;
	Construct rank-$p$ approximating basis\; \nonl $\mW_{fr} = \bmat{\m w_{f_1}\;,\m w_{f_2}\;,\cdots, \m w_{f_p}}$\;
	Choose the first index: $[\rho$, $i_1]$ $\hspace{-0.00cm}=\hspace{-0.00cm}\mr{max}(\m w_{f_1})$\;
	Construct first measurement matrix\; \nonl $\mP_{M_1}\hspace{-0.00cm} =\hspace{-0.00cm}  \m e_{i_1},\, \mW_{fr} \hspace{-0.00cm}=\hspace{-0.00cm} [\m w_{f_1}]$\;
	\For{$j = 2:p$}{
		calculate $\m c$ using, $\mP_M^\top\mW_{fr}\m c = \mP_M\m w_{f_j}$\;
		compute residual, $\m d = \m w_{f_j}-\mW_{fr}\m c$\;
		update $\mP_M$ and $\mW_{fr}$ as follows:\;
		$[\rho, i_j] =\mr{max}(\m d)$\;
		$\mW_{fr} =[\mW_{fr}, \m w_{f_j}], \, \mP_M = [\mP_M, \m e_{i_j}]$}
	Calculate ${\m f}_r(\cdot)$ \eqref{eq:nonlin_apprx} and $\mE_r$, $\mA_r$, $\mB_{ur}$, $\mB_{wr}$ using \eqref{eq:ROM_Mat} 
\end{algorithm}
Notice that in case $t<<n_d$ and/or $t<<n_a$, then to simplify complexity and save computational time one can take eigenvalue decomposition (ED) of $\mX_d^\top\mX_d \in\mbb{R}^{t\times t}$ and/or $\mX_a^\top\mX_a \in\mbb{R}^{t\times t}$  and then design $\mW$ as follow: For $\mX_d^\top\mX_d$ applying ED we get $\mX_d^\top\mX_d\mV_d= \m V_d\m \lambda_d$, where $\mV_d$ encapsulates the corresponding eigenvectors and $\m\lambda_d$ contains the eigenvalues. Similarly, for $\mX_a^\top\mX_a$ we have $\mX_a\mX_a^\top\mV_a = \m V_a\m \lambda_a$ with matrices $\m V_a \in\mbb{R}^{t\times t}$ and $\m \lambda_a \in\mbb{R}^{t\times t}$ lumping the eigenvectors and eigenvalues, respectively. Then, we design the matrices $\mW_{d} = \mX_d\mV_d\m\lambda_d^{\frac{-1}{2}}$ and $\mW_{a} =  \mX_a\mV_a\m\lambda_a^{\frac{-1}{2}}$ and finally plugging it in \eqref{eq: W mat} gives us the final transformation matrix $\mW$. The truncation matrix $\mT$ can be designed similarly to Eq. \eqref{eq:T equation}.

The final step in the proposed SP-POD is the handling of the nonlinearity. Note that, using the designed $\mT$ and $\mW$ we can express the corresponding ROM of model \eqref{eq:final_NDAE} as given in Eq. \eqref{eq:ROM_initial}. However, the computational complexity of mapping the nonlinear function $\mW_L{\m f}\left({\m W_R \tilde{\m x}},{\m w},{\m u} \right)$ still depends on the dimension of full state vector $\m x$ as:
\begin{align}
	\underbrace{\mW_L}_{r\times n}\underbrace{{\m f}\left({\m W_R \tilde{\m x}},{\m u},{\m w} \right)}_{n\times 1}.
\end{align}
Therefore, we reduce the nonlinearity using the discrete empirical interpolation method (DEIM) \cite{brunton2019data, 9637933, chaturantabut2009discrete}. The primary goal is to handle the nonlinear terms efficiently within a reduced-dimensional space instead of the full-dimensional space $\mathbb{R}^{n}$. DEIM achieves this by selecting a subset of spatial locations (interpolation points) where the nonlinear function is evaluated. By measuring specific points in the state space rather than the entire set of state variables, DEIM approximates the nonlinear term through interpolation around these selected points, this approximation can be expressed as \cite{brunton2019data}:
\begin{align}\label{eq:DIEM1}
\underbrace{\mW_L\mW_{fr}}_{r\times p}\underbrace{{\m f}_r\left( \cdot \right)}_{p\times 1}.
\end{align}
% \begin{align}\label{eq:DIEM1}
% \underbrace{\mW_L\mW_{fr}}_{r\times p}\underbrace{{\m f}_r\left( \cdot \right)}_{p\times 1}.
% \end{align}
The main aim is to project the original nonlinearity ${\m f}\left({\m W_R \tilde{\m x}},{\m u},{\m w} \right)$ onto $\mW_{fr}$ such that ${\m f}\left({\m W_R \tilde{\m x}},{\m u},{\m w} \right) \approx \mW_{fr}{\m f}_r\left(\cdot \right)$ with $\mW_L\mW_{fr}$ being precomputed offline. 

To carry out such approximation for the nonlinear term  we start by storing the snapshots of the  ${\m f}\left({{\m x}},{\m u},{\m w} \right)$ in a matrix given as:
\begin{align}\label{eq:f data}
	\m X_f &= \bmat{
		\mid & \mid & & \mid & \mid\\
		{\m f}_{0} & {\m f}_{1} & \cdots & {\m f}_{{t-1}} & {\m f}_{t}\\
		\mid & \mid & & \mid & \mid}. 
\end{align}
Note that, to construct matrix $\m X_f$, the function $\m f(\m x, \m u, \m w)$ is evaluated for each time snapshot of $\m x_d$, $\m x_a$ (stored in data matrices $\mX_d$ and $\mX_a$), and the corresponding $\m u$ and $\m w$. For example $\m f_0 = \m f(\m x_0, \m u_0, \m w_0), \quad \m f_1 = \m f(\m x_1, \m u_1, \m w_1)$, and so on. Then we take SVD of this snapshot matrix, $\m X_f = \mW_f\m\Sigma_f\m\Lambda_f$ with matrix $\mW_f = \bmat{\m w_{f_1}\;,\m w_{f_2}\;,\cdots, \m w_{f_n}} \in\mbb{R}^{n\times n}$ containing the left singular vectors, matrix $\m\Lambda_f \in\mbb{R}^{t\times n}$ lumping the right singular vectors, and $\m\Sigma_f \in\mbb{R}^{n\times n}$ being the diagonal matrix containing the HSVs. Next, we design $\mW_{fr} = \bmat{\m w_{f_1}\;,\m w_{f_2}\;,\cdots, \m w_{f_p}} \in\mbb{R}^{n\times p}$ as the first $p$ columns of $\mW_f$. Finally, we design a binary measurement matrix $\mP_M$ that selects optimal points in the reduced subspace $\mW_{fr}$ so that nonlinearity can be reconstructed using the selected points efficiently. To construct such $\mP_M$ we utilize residual-based greedy technique \cite{brunton2019data, chaturantabut2009discrete}, which essentially puts a measurement (or a $1$ in matrix $\mP_M$) where residual/error is maximum (highlighting measurement point is required).  These points are chosen to maximize the approximation accuracy. The greedy algorithm proceeds as follows:

The algorithm selects the first measurement location based on the maximum value in the first mode, $\m w_{f_1}$. Selecting the maximum value as the first measurement point ensures that the initial point captures a critical aspect of the nonlinear term $\m f(\cdot)$. After establishing the first measurement point, the algorithm iterates to select additional points.  In each iteration, it computes the projection of the current modes onto the next ones as:
\begin{align}
\mP_M^\top\mW_{fr}\m c = \mP_M\m w_{f_j}
\end{align}
where $\m c$ denotes the projection of the current modes contained in  $\mW_{fr}$ onto the next mode $\m w_{f_j}$. Then, the residual is computed as:
\begin{align}
	\m{d} = \m w_{f_j}-\mW_{fr}\m c
\end{align}
and the next measurement point is selected where the value of $\m{d}$ is maximum. By selecting points with the maximum residual, the algorithm ensures that each new measurement location adds the most significant new information about the nonlinear term $\m f(\cdot)$. After completion of the iterations, the approximation to the nonlinearity can be expressed as:
\begin{align}\label{eq:nonlin_apprx}
    {\m f}_r(\cdot) = \mW_{fr}(\mP_M^\top\mW_{fr})^{-1} {\m f}\left({\mP_M^\top\m W_R \tilde{\m x}},{\m u},{\m w} \right)
\end{align}

We note that, if the approximation of the nonlinearity in the projected basis using the presented DIEM is insufficient, increasing the number of interpolation points can enhance accuracy. Also, if the computational cost of mapping the nonlinearity is low (such as form most of power system models), the DIEM step in the proposed algorithm may be omitted entirely. Additionally, other greedy algorithms such as Q-DIEM or EIM can also be utilized as effective substitutes for DIEM to efficiently reduce the computational complexity of handling the nonlinearity \cite{brunton2019data}.

That said, the final ROM  can be written as follows:
\begin{subequations}\label{eq:ROM_final}
	\begin{align}
		\hspace{-0.2cm}\m E_r\dot{{\m z}} &= \m A_r{\m z} + \m B_{ur}{\m u} + \mW_L{\m f}_r\left(\cdot\right) + \m B_{wr} \m w\\
		\m y &= \mC_r\m z.
	\end{align}
\end{subequations}
The overall proposed SP-POD-based MOR algorithm is summarized in Algorithm \ref{alg:Algorithm 1}. In the following section, we present the structure-preserving balanced POD-based (SP-BPOD) MOR technique for the complete NDAE representation of power systems.
\section{Structure-Preserving BPOD MOR}\label{sec:mor2}
\vspace{-0.0cm}
The SP-POD presented in the previous section performs model reduction based on the energy contents (or modes) in the time-domain simulation data. While SP-POD is effective in capturing the dominant behaviors of a system, it primarily focuses on the energy content without directly considering the impact of these modes on the system's controllability or observability. This limitation can be critical in control applications, as modes with lower energy levels might still substantially impact how the system responds to controls and how well it can be monitored or observed \cite{brunton2019data}. 
Therefore, here we also propose a structure-preserving balanced POD-based MOR approach for the power system NDAE dynamics. This approach involves reordering the system states based on a balance criterion that accounts for controllability and observability. Thus, only modes that are both highly controllable and highly observable are
retained while the rest are truncated, making balanced models ideal for control applications.

To design balanced models, we need to compute the controllability and observability Gramians. Now, it is well-known that solving the Lyapunov equations to compute Gramians for a much larger system model can be very challenging \cite{freitas2008gramian}. Also, in our case, we have to solve the \textit{generalized} Lyapunov equation (since the considered system is DAE and not ODE) which is even much harder to solve and becomes numerically intractable \cite{Romijn}. Furthermore, using Lyapunov equations to compute Gramians only considers the system's linear part (through the system matrices $\mA$, $\mB_u$, etc.) and ignores the accompanying nonlinear function $\m f(\cdot)$. Thus, the Gramians may only be valid in the vicinity of the equilibrium point.
Given these challenges, in the literature, the idea of empirical covariance matrices have been introduced in \cite{lall2002subspace, hahn2002balancing}, which approximate the system Gramians from system impulse responses. It has been shown that for linear time-invariant systems, empirical covariances are exactly equal to the usual Gramians derived from system matrices \cite{lall2002subspace}. In the following sections, we briefly introduce these covariance matrices and further details can be found in \cite{lall2002subspace, hahn2002balancing}.
 
\vspace{-0.1cm}
\subsection{Empirical Controllability Covariance}
\vspace{-0.051cm}
To state the empirical controllability covariance we first define the following sets:
\begin{align*}
	& \mathcal{T}^c=\left\{\boldsymbol{T}_1^c,\cdots, \boldsymbol{T}_q^c;\boldsymbol{T}_l^c \in \mathbb{R}^{n_u \times n_u}, \boldsymbol{T}_l^{c \top} \boldsymbol{T}_l^c=\boldsymbol{I}, l=1,\cdots, q\right\} \\
	& \mathcal{M}^c=\left\{c_1^c, \cdots, c_s^c;\;c_m^c \in \mathbb{R},\; c_m^c>0,\; m=1, \cdots, s\right\} \\
	& \mathcal{E}^c=\left\{\boldsymbol{e}_1^c,\cdots, \boldsymbol{e}^c_{n_u}; \text { standard unit vectors in } \mathbb{R}^{n_u}\right\}
\end{align*}
where $\mathcal{T}^c$ represents the set of excitation direction matrices and it contains $q$ orthogonal excitation matrices, each of size ${n_u \times n_u}$, the set $\mathcal{M}^c$ denotes the set of excitation magnitudes and it comprises of $s$ positive real numbers, each representing a different magnitude of excitation to apply along the directions specified in $\mathcal{T}^c$, and the set  $\mathcal{E}^c$ defines the control input to be excited. Using the above sets, perturbations in the control input for each time step $k$ can be written as $\boldsymbol{u}(k)=c_m \boldsymbol{T}_l^c \boldsymbol{e}_i \mathrm{u}(k)+ \boldsymbol{u}_0(0)$ with $c_m$ specifying the magnitude, $\m T_l^c\m e_i$ expressing the direction, and $\mathrm{u}(k)$ representing the temporal shape of the perturbation.

The empirical controllability covariance can then be expressed as follows:
\begin{align}\label{eq:Wc}
	\boldsymbol{G}_c=\sum_{i=1}^{n_u} \sum_{l=1}^q \sum_{m=1}^s \frac{1}{q s c_m^2} \sum_{k=0}^{K} \boldsymbol{\Psi}^{i l m}(k) \Delta t(k)
\end{align}
where $K$ represents the number of points chosen to approximate the covariance matrix, the notation $\boldsymbol{\Psi}^{i l m}(k)=\left(\boldsymbol{x}^{i l m}(k)-\right.$ $\left.\boldsymbol{x}_0^{i l m}\right)\left(\boldsymbol{x}^{i l m}(k)-\boldsymbol{x}_0^{i l m}\right)^{\top}$ and it quantifies the change in the system's state from its initial state.  The vector $\boldsymbol{x}_0^{i l m}$ represents the steady-state of the system while  $\boldsymbol{x}^{i l m}(k)$ represents the state of the system at time-step $k$ influenced by an input $\m{u}(k)$

\vspace{-0.03cm}
\subsection{Empirical Observability Covariance}
\vspace{-0.01cm}
Similarly to as done previously, we define the following sets for empirical observability covariance:
\begin{align*}
	& \mathcal{T}^o=\left\{\boldsymbol{T}_1^o,\cdots, \boldsymbol{T}_q^o;\boldsymbol{T}_l^o \in \mathbb{R}^{n \times n}, \boldsymbol{T}_l^{o \top} \boldsymbol{T}_l^o=\boldsymbol{I}_n, l=1,\cdots, q\right\} \\
	& \mathcal{M}^o=\left\{c_1^o, \cdots, c_s^o ;\; c_m^o \in \mathbb{R},\;c_m^o>0,\;m=1,\cdots, s\right\} \\
	& \mathcal{E}^o=\left\{\boldsymbol{e}_1^o,\cdots, \boldsymbol{e}^o_{n}; \text { standard unit vectors in } \mathbb{R}^{n}\right\}
\end{align*}
where set $\mathcal{T}^o$ represents the state excitation directions with total $q$ orthogonal excitation matrices, set $\mathcal{M}^o$ defines the excitation magnitudes, and $\mathcal{E}^o$ defines the state to be excited. Then, we can define the initial condition perturbation vector as:  $\m{x}(0)=$ $c_m \mT_l^o \m e_i+\boldsymbol{x}_0$, where $c_m$ dictates the perturbation magnitude and $\mT_l^o \m e_i$ decides the perturbation direction.

The empirical observability covariances is then expressed as follows:
\begin{align}\label{eq:Wo}
\boldsymbol{G}_o=\sum_{l=1}^q \sum_{m=1}^s \frac{1}{q s c_m^2} \sum_{k=0}^{K} \boldsymbol{T}_l^o \boldsymbol{\Psi}^{l m}(k) \boldsymbol{T}_l^{o \top} \Delta t(k)
\end{align}
where $\m{\Psi}^{l m} (k)\in \mathbb{R}^{n \times n}$ with $\Psi_{i j}^{l m}(k)=\left(\m y^{i l m}(k)-\right.$ $\left.\m y_0^{i l m}\right)^{\top}\left(\m y^{j l m}(k)-\m y_0^{j l m}\right)$ representing the change in the system output from its equilibrium $\m y_0^{i l m}$ when influenced by change in the system initial condition given by $\m{x}(0)$ as defined above.

%where $\m \Psi_{i j}^{l m}(k)=\left(\m y^{i l m}(k)-\right.$ $\left.\m y_0^{i l m}\right)^{\top}\left(\m y^{j l m}(k)-\m y_0^{j l m}\right)$ lumps the system responses to the initial conditions $\boldsymbol{x}(0)=$ $c_m T_l^o \m e_i+\boldsymbol{x}_0$ with $\m y_0^{i l m}$ representing the output of the system at the equilibrium point $\m x_0$. 

\vspace{-0.01cm}
\subsection{Empirical Balanced Model Synthesis}
\vspace{-0.01cm}
 Given the empirical covariances, we now have the necessary tools to transform the NDAE power system \eqref{eq:final_NDAE} to other coordinates where the system states are ordered and balanced. To do that, the system needs to be scaled first. This ensures that states changing by orders of magnitude are appropriately accounted for in their significance to the system's dynamics, compared to states with minimal changes. Therefore, we define the following scaled/normalized vectors:
 \begin{align}
 	{\m x}_{s} =  \mS^{-1}_x \m x, \;\;\; {\m u}_{s} =  \mS^{-1}_u \m u, \;\;\; {\m w}_{s} =  \mS^{-1}_w \m w 
 \end{align}
where  ${\mS_x} = \mr{diag}(\m x_0)$, ${\mS_u} = \mr{diag}(\m u_0)$, and ${\mS_w} = \mr{diag}(\m w_0)$ with $\m x_0$,  $\m u_0$, and  $\m w_0$ representing the steady-state values of these vectors. Then, the scaled representation of \eqref{eq:final_NDAE} can be written as follows:
\begin{subequations}\label{eq:final_NDAE_scalled}
 	\begin{align}
 		\hspace{-0.3cm}	\m E\dot{\m x}_{s} &\hspace{-0.05cm}= \hspace{-0.05cm}\mS^{-1}_x{\m A} \mS_x{\m x}_{s} \hspace{-0.05cm} + \hspace{-0.05cm}\mS^{-1}_x\hspace{-0.09cm}{{{\m  B}}_u} \mS_u{\m u}_{s}\hspace{-0.05cm}   +\hspace{-0.05cm}   \mS^{-1}_x\hspace{-0.09cm}{\m f}\hspace{-0.05cm} +\hspace{-0.05cm}  \mS^{-1}_x\hspace{-0.09cm}{\m B}_w \mS_w\m w_{s}\\
 		\m y_{s} &= \m C\mS_x \m x_{s}.
 	\end{align}
 \end{subequations}
where ${\m f} = {\m f}\left({\m x_{s}},{\m u}_{s},{\m w}_{s} \right)$. From now on, for the sake of simplicity, with a little abuse of notation, we consider $\m x_{s} = \m x$, $\m u_{s} = \m u$, and $\m w_{s} = \m w$. Similarly, let $\mA =  \mS^{-1}_x\hspace{-0.09cm}{\m A} \mS_x$, $\mB_u =  \mS^{-1}_x\hspace{-0.09cm}{\m B_u} \mS_u$, $\mB_w =  \mS^{-1}_x\hspace{-0.07cm}{\m B_w} \mS_w$, and $\mS^{-1}_x\hspace{-0.09cm}{\m f}\left({\m x_{s}},{\m u}_{s},{\m w}_{s} \right) = {\m f}\left({\m x},{\m u},{\m w} \right)$.

The next step is to perform coordinate transformation such as $\tilde{\m x}= \mW {\m x}$ so that the system is balanced.  Notice that, for ODE systems using empirical covariance matrices, computing coordinate transformation matrix $\mW$ is straightforward and well-documented, often involving Cholesky factorization or similar techniques--see \cite{lall2002subspace,hahn2002balancing}. However, for NDAE systems, the presence of algebraic equations (which do not exhibit dynamic behavior and represent static constraints) complicates the application of these techniques. The NDAE system evolves in a subspace defined by the differential equations, while algebraic equations restrict this evolution without contributing to the system's dynamics \cite{Romijn}. Hence, motivated by \cite{sun2005reduction} we propose a two-step approach to design a balanced model. In the first step, we perform balancing for the dynamic variables and then in the second step we perform coordinate transformation for algebraic variables so that they can also be truncated, the details are given in the subsequent sections.
\vspace{-0.4cm} 	
\subsection{Balancing Dynamic Variables}
\vspace{-0.1cm}
To perform balancing for the dynamic variables, we first need to compute their covariance matrices. With that in mind, to design controllability covariance the NDAE system \eqref{eq:final_NDAE} can be excited by perturbations in the control inputs, and state trajectories can be generated. From these trajectories, a covariance matrix can be computed using the Eq. \eqref{eq:Wc}. This covariance matrix includes the controllability covariance matrix for the states governed by differential equations. Meaning the designed $\m G_c$ can be decomposed as follows:
\begin{align}\label{eq:WC dcomposed}
	\m G_c = \bmat{\m G_{c_{11}}& \m G_{c_{12}}\\\m G_{c_{21}}& \m G_{c_{22}}} 
\end{align}
with $\m G_{c_{11}} \in \mathbb{R}^{n_d \times n_d}$, $\m G_{c_{12}} \in \mathbb{R}^{n_d \times n_a}$, $\m G_{c_{21}} \in \mathbb{R}^{n_a \times n_d}$, and $\m G_{c_{22}} \in \mathbb{R}^{n_a \times n_a}$.  In \eqref{eq:WC dcomposed} $\m G_{c_{11}}$ is the symmetric positive-definite controllability covariance for the dynamic variables and similarly $\m G_{c_{22}}$ is the covariance matrix for the algebraic variables. 
Notice that, $\m G_{c_{22}}$ does not represent controllability in the traditional sense. Instead, it indicates correlations among algebraic variables and can be useful for reducing their number (which is discussed in detail in Sec. \ref{sec:coordinate_algebric-E}).

Similarly, observability covariance matrix can be obtained by introducing systematic perturbations in the system's initial conditions $\m x_0$ as explained in the previous section. However, limited information regarding the system's observability can be gathered. This is due to the fact that in a regular DAE system, there are only as many degrees of freedom for selecting consistent initial conditions as there are dynamic variables \cite{hahn2003controllability,sun2005reduction}. As such, once each dynamic variable has been perturbed independently, no additional information regarding the system's observability can be obtained by perturbing the algebraic states \cite{hahn2003controllability}. In reality, the perturbation of algebraic variables only produces an initial condition that locally signifies a linear combination of the perturbations that were previously applied to the differential variables. Therefore, by solely perturbing dynamic states, the covariance matrix $\m G_{o_{11}} \in \mathbb{R}^{n_d \times n_d}$ is determined using Eq. \eqref{eq:Wo}.

% Similarly,  the observability covariance matrix can be determined by adding systematic perturbation in the system's initial conditions $\m x_0$ and using Eq. \eqref{eq:Wo}. However, notice that only limited information regarding system observability can be extracted. This is because, for a regular DAE, there are only as many degrees of freedom for choosing consistent initial conditions as it has dynamic variables \cite{hahn2003controllability,sun2005reduction}. Hence, once each dynamic variable has been perturbed independently, no additional information about the system's observability can be obtained by perturbing the algebraic variables. In fact, perturbing the algebraic states only yields an initial condition that locally represents a linear combination of the perturbations previously applied to the differential variables. This means that after the initial exploration of the differential equations, the manipulation of algebraic variables does not provide new knowledge about the system's behavior \cite{hahn2003controllability,sun2005reduction}. Therefore, by only perturbing dynamic states the covariance matrix $\m G_{o_{11}} \in \mathbb{R}^{n_d \times n_d}$ has been determined using Eq. \eqref{eq:Wo}.

That being said, using $\m G_{c_{11}}$ and $\m G_{o_{11}}$ we can balance the dynamic variables. The main objective is to find a coordinate transformation matrix that can make $\m G_{c_{11}}$ and $\m G_{o_{11}}$ diagonal and equal in a new state coordinates that are both observable and controllable. To find such transformation, a technique has been proposed in \cite{hahn2002balancing} which decouples the system into controllable/uncontrollable as well as observable/unobservable components similar to Kalman decomposition. The overall procedure involves four main steps and are given as follows:
\begin{enumerate}
	\item The first step is to isolate controllable states by transforming the controllability covariance matrix $\m G_{c_{11}}$ into a block diagonal form. This is done by applying a Schur decomposition to $\m G_{c_{11}}$ using a unitary transformation matrix $\mT_1$, resulting in a block diagonal matrix that highlights the rank (and hence the controllability) of the system. This transformation is given as follows:
	\begin{align}\label{eq:T_1}
		\m T_1 \m{G}_{c_{11}} \mT_1^\top= \bmat{			\mI & \m0 \\
			\m0 & \m0}
	\end{align}
where the identity matrix $\mI$ represents the fully controllable states. 
\item In the second step, the transformation $\m T_1$ computed in the previous step is applied to the observability covariance $\m G_{o_{11}}$ as:
\begin{align}
	 	\left(\mT_1^\top\right)^{-1} {\m G_{o_{11}}}\mT_1^{-1}=\bmat{\tilde{\mG}_{1} & \tilde{\mG}_{2} \\
	 		\tilde{\mG}_{3} & \tilde{
	 			\mG}_{4}}.
\end{align}
Then a Schur decomposition of the upper block $\tilde{\mG}_{1}$ is carried out to isolate observable states as:
\begin{align}
	 \mL_1 \tilde{\mG}_{1} \mL_1^\top=\bmat{		\m\Gamma_1{ }^2 & \m0 \\
	 	\m0 & \m0}
\end{align}
The resulting unitary matrix $\mL_1$ from this decomposition forms the basis of the second transformation matrix given as:
\begin{align}\label{eq:T_2}
\left(\mT_2^{\top}\right)^{-1}=\bmat{\mL_1 & \m0 \\
		\m0 & \mI}
\end{align}

\item In the third step, the combined transformations 
$\mT_1$ and $\mT_2$ are applied to the original observability covariance matrix $\m G_{o_{11}}$ to  further isolate observable and controllable states and to construct the third transformation matrix $\m T_3$ as follows:
\begin{align*}
	 	\left(\m T_2^{\top}\right)^{-1}\left(\mT_1^{\top}\right)^{-1} \mathbf{G}_{o_{11}}\mT_1^{-1}\mT_2^{-1}= \bmat{	\m\Gamma_1{ }^2 & \m0 & \hat{\mG}_{2} \\
	 		\m0 & \m0 & \m0 \\
	 		\hat{\mG}_{2}{ }^{\top} & \m0 & \hat{\mG}_{4}}
 		\end{align*}
 and $\mT_3$ is given as:
\begin{align}\label{eq:T_3}		
\left(\mT_3^{\top}\right)^{-1}=\bmat{\mI & \m0 & \m0 \\
		\m0 & \mI & \m0 \\
		-\hat{\mG}_{2}{ }^{\top} \m\Gamma_1{ }^{-2} & \m0 & \mI}.
\end{align}
\item In the final fourth step, we apply a sequence of transformations ($\mT_1$ through $\mT_3$) to the observability covariance matrix. Then, we perform Schur decomposition on the element of the last column and row of the resultant matrix to construct the final transformation matrix $\mT_4$ as follows:
\begin{align*}
\left(\mT_3^{\top}\right)^{-1}\left(\mT_2^{\top}\right)^{-1}\left(\mT_1^{\top}\right)^{-1} \mathbf{G}_{o_{11}}\mT_1^{-1}\mT_2^{-1}\mT_3^{-1}\\=
\bmat{ \m\Gamma_1{ }^2 & \m0 & \m0 \\
	\m0 & \m0 & \m0 \\
	\m0 & \m0 & \tilde{\mG}_{4}-\hat{\mG}_{2}{ }^{\top} \m\Gamma_1{ }^{-2} \hat{\mG}_{2}}
\end{align*}
and
\begin{align*}
\mL_2( \tilde{\mG}_{4}-\hat{\mG}_{2}{ }^{\top} \m\Sigma_1{ }^{-2} \hat{\mG}_{2})\mL_2^\top = \bmat{ 			\m\Gamma_3 & \m0 \\
	\m0 & \m0}
 \end{align*}
\begin{align}\label{eq:T_4}
\left(\mT_4^{\top}\right)^{-1}= \bmat{	\m\Gamma_1^{-1 / 2} & \m0 & \m0 \\
	\m0 & \mI & \m0 \\
	\m0 & \m0 & \mL_2}.
\end{align}
\end{enumerate}
Finally, the complete coordinate transformation matrix $\m W_d$ that leads to the balanced forms of $\m G_{c_{11}}$ and $\m G_{o_{11}}$ can be constructed by multiplying all the individual transformations ($\mT_1$ through $\mT_4$), as follows:
\begin{align}\label{eq:Wd}
	\mW_d = \mT_1\mT_2\mT_3\mT_4
\end{align}
and the corresponding balanced covariance matrices are given as:
\begin{align}
	\mW_d \m G_{c_{11}} \mW_d^{\top} &= \bmat{		\m \Gamma_1 & \m0 & \m0 & \m0 \\
		\m0 & \mI & \m0 & \m0 \\
		\m0 & \m0 & \m0 & \m0 \\
		\m0 & \m0 & \m0 & \m0} \label{eq:sigma1}\\
	\left(\mW_d^{-1}\right)^{\top} \m G_{o_{11}}(\mW_d)^{-1} &= \bmat{	\m\Gamma_1 & \m0 & \m0 & \m0 \\
		\m0 & \m0 & \m0 & \m0 \\
		\m0 & \m0 & \m\Gamma_3 & \m0 \\
		\m0 & \m0 & \m0 & \m0} \label{eq:sigma3}.
\end{align}
The final transformation matrix $\mW_d$ decomposes the dynamic variables into four separate categories, states that are $(1)$ both controllable and observable, $(2)$ controllable but not observable, $(3)$ observable but not controllable, and $(4)$ neither observable nor controllable. In particular, the diagonal matrix $\m\Gamma_1$ (with diagonal entries representing the HSVs) signifies the states that are both controllable and observable, the identity matrix $\mI$ depicts the states that are only controllable, the diagonal matrix $\m\Gamma_3$ denotes the states that are observable but not controllable, while zeros represent the states that are neither controllable nor observable.  The above-balanced form facilitates system simplification by eliminating states that contribute little to the system's dynamic behavior.
%This decomposition is akin to the Kalman canonical form, which separates a system into its most critical (controllable and observable) and least critical (uncontrollable or unobservable) components.
\vspace{-0.24cm}
\begin{algorithm}[h]\label{alg:Algorithm 2}
	\caption{\text{SP-BPOD for power system NDAE models}}
	\DontPrintSemicolon 
    \textbf{Input:} NDAE~\eqref{eq:final_NDAE_scalled} parameters $\mA$, $\mB_u$, $\mB_{w}$, $\mE$, $\m f(\cdot)$, and $\m x_0$\;
    \textbf{Output:} ROM parameters  $\mA_r$, $\mB_{ur}$, $\mB_{wr}$, $\mE_r$, and $\m f_r(\cdot)$\;
	Compute matrices $\mG_{c_{11}}$  and  $\mG_{o_{11}}$ using \eqref{eq:Wc} and \eqref{eq:Wo}\;	
	Compute $\mT_1$ \eqref{eq:T_1}, $\mT_2$ \eqref{eq:T_2}, $\mT_3$ \eqref{eq:T_3}, and $\mT_4$ \eqref{eq:T_4}\;
	Construct $\mW_d = \mT_1\mT_2\mT_3\mT_4$ as in \eqref{eq:Wd}\;
	Perform SVD of $\mG_{c_{22}}$ as:
	$\m G_{c_{22}} = \m W_{gc}\m\Sigma_{gc}\m \Lambda_{gc}$ \eqref{eq:svd of Gc22}\;
	Select $\mW_a = \m W_{gc}$ \;
	Compute $\mW = \mr{blkdiag}(\mW_d,\; \mW_a)$ as in \eqref{eq: W mat}\;
	Design $\mT_a$, $\mT_d$ by examining HSVs in  $\m\Sigma_{gc}$ \eqref{eq:svd of Gc22} and $\m\Gamma_1$ \eqref{eq:sigma1}\;
	Design matrix $\mT = \mr{blkdiag}(\mT_d,\; \mT_a)$ as in \eqref{eq:T equation}\;
	Design $\mW_L = \mW\mT$ and  $\mW_R = \mW_L^{-1}$\;
	Construct $\mX_f$ \eqref{eq:f data} and SVD it as: 
	 $\m X_f = \mW_f\m\Sigma_f\m\Lambda_f$\;
	Start greedy algorithm to design $\mP_M$\;
	Construct rank-$p$ approximating basis\; \nonl $\mW_{fr} = \bmat{\m w_{f_1}\;,\m w_{f_2}\;,\cdots, \m w_{f_p}}$\;
	Choose the first index: $[\rho$, $i_1]$ $\hspace{-0.05cm}=\hspace{-0.05cm}\mr{max}(\m w_{f_1})$\;
	Initialize measurement matrix\; \nonl $\mP_{M_1}\hspace{-0.08cm} =\hspace{-0.05cm}  \m e_{i_1},\, \mW_{fr} \hspace{-0.05cm}=\hspace{-0.05cm} [\m w_{f_1}]$\;
	\For{$j = 2:p$}{
		calculate $\m c$ using, $\mP_M^\top\mW_{fr}\m c = \mP_M\m w_{f_j}$\;
		compute residual, $\m d = \m w_{f_j}-\mW_{fp}\m c$\;
		update $\mP_M$ and $\mW_{fr}$ as follows:\;
		$[\rho, i_j] =\mr{max}(\m d)$\;
		$\mW_{fr} =[\mW_{fr}, \m w_{f_j}], \, \mP_M = [\mP_M, \m e_{i_j}]$}
	Calculate ${\m f}_r(\cdot)$ \eqref{eq:nonlin_apprx} and $\mE_r$, $\mA_r$, $\mB_{ur}$, $\mB_{wr}$ using \eqref{eq:ROM_Mat} 
\end{algorithm}
\vspace{-0.1cm}
\subsection{Coordinate Transformation of Algebraic Variables}\label{sec:coordinate_algebric-E}
\vspace{-0.1cm}
In the above section we performed coordinate transformation for dynamic states such that they are balanced and hierarchically ordered and thus suitable for reduction. Here, we perform transformation for the algebraic variables so that they can also be ordered and later on truncated. This can simply be done by taking the SVD of covariance matrix $\m G_{c_{22}}$ (which is computed in Eq. \ref{eq:WC dcomposed}) as follows:
\begin{align}\label{eq:svd of Gc22}
	\m G_{c_{22}} = \m W_{gc}\m\Sigma_{gc}\m \Lambda_{gc}
\end{align}
where $\m\Sigma_{gc}$ is a diagonal matrix and contains the HSVs in descending order while matrices $\m\Lambda_{gc}$ and $\m W_{gc}$ contain the right and left singular vectors, respectively. The columns of $\m W_{gc}$ are ordered hierarchically from most dominant to least and are referred to as the modes of $\m G_{c_{22}}$. Hence, the coordinate transformation matrix $\m W_a$ for the algebraic variables can be set to be equal to $\m W_{gc}$, i.e. $\m W_a = \m W_{gc}$.

The final non-singular coordinate transformation matrix $\mW$ and the truncation matrix $\mT$ for the SP-BPOD can be expressed similarly as in \eqref{eq: W mat} and \eqref{eq:T equation}.  While the dimensions of identity matrices $\mI_{dr}$ and $\mI_{ar}$ here can be designed by examining the magnitude of HSVs in $\m \Gamma_1$ and $\m\Sigma_{gc}$, respectively. The corresponding matrices $\mW_R$ and $\mW_L$ for the SP-BPOD can then constructed as  $\mW_L = \m T\m W$ and $\m W_R =  \mW_L^{-1}$.

Now as discussed in the previous section, mapping the nonlinearity using $\mW_R$ and $\mW_L$ as $\mW_L{\m f}\left({\m W_R \tilde{\m x}},{\m w},{\m u} \right)$ still depends on the dimension of full state vector $\m x$ and thus can be computationally expensive. Then again, one can use the DEIM-based hyper-reduction approach to measure specific points in the state-space and then efficiently interpolate the nonlinearity around the selected points (as done in Eq. \ref{eq:DIEM1} of SP-POD-based MOR technique). Having said that, the overall proposed SP-BPOD-based MOR algorithm for the complete NDAE power system model is summarized in Algorithm \ref{alg:Algorithm 2}. It is worth mentioning here that in both the proposed SP-POD and SP-BPOD model reduction techniques, the final reduced order model is guaranteed to be an NDAE. In both proposed techniques, the number of dynamic states in the ROM is determined by the dimensions of $\mI_{dr}$; similarly, the number of algebraic variables is controlled by the dimension of user-defined identity matrix $\mI_{ar}$. 

\vspace{-0.1cm}
\section{Case Studies}\label{sec:case studies}
\vspace{-0.05cm}
To assess the effectiveness of the proposed methods we perform thorough simulation studies on various power system models, namely, the modified IEEE 39-bus and the 2000-bus Texas networks \cite{Texas_system}.  The details about these test systems are given in the below sections and Appendix \ref{appndix:ninth Gen_dynamics}.  All the numerical simulations are performed on MATLAB R$2023a$ running on a personal laptop with an Intel-$i9$ processor. The NDAE power system models are simulated using MATLAB index-1 DAEs solver \texttt{ode15s}. The system volt-ampere base is chosen to be $S_b = 100\mr{MVA}$ while the frequency base is selected as $w_b = 120\pi\mr{rad/s}$. To carry out the time-domain simulations the system's initial conditions are determined using power flow studies carried out in MATPOWER.  

To implement the proposed SP-POD-based MOR technique, the snapshot data matrices $\mX_d$, $\mX_a$, and $\mX_f$ need to be computed. This is done by carrying out time domain simulations under step disturbance in overall system load demand as follows.  Initially, the system operates under steady-state conditions, meaning load demand is exactly equal to generation and thus there are no transients. Then, right at $t>0$ an abrupt disturbance in load demand is applied to the system as: $\mP^e_d + j\mQ^e_d = (\mI +\Delta_d)(\mP^0_d + j\mQ^0_d)$,
where $\mP^0_d$ and $\mQ^0_d$ represent the initial active/reactive load demand and $\mP^e_d$ and $\mQ^e_d$ are their respective values after the disturbances. The parameter $\Delta_d$ denotes the severity of the disturbances. The overall simulation time period is set to be $20$sec and the system transients dynamic and algebraic states data is saved in matrices $\mX_d$, $\mX_a$, and $\mX_f$. These snapshot data is then used in SP-POD Algorithm  \ref{alg:Algorithm 1} to design the corresponding ROM.

For the SP-BPOD-based MOR algorithm the empirical covariances matrices $\mG_{c}$ and $\mG_{o_{11}}$ are computed for the scaled system \eqref{eq:final_NDAE_scalled} over the time period $[0, 5s]$ with $\Delta t(k)$ set to be $0.01s$. To design the covariances matrices using Eqs. \eqref{eq:Wc} and \eqref{eq:Wo} systematic perturbations in the control inputs $\m u$ or initial conditions $\m x_0$ are added right at $t>0$ by defining the following sets:
\begin{align}\label{eq:T set}
	\mathcal{T}^c=\left\{\m{I}_{n_u \times n_u},-\m I_{n_u \times n_u}\right\},\;\;\;
	\mathcal{T}^o=\left\{\m{I}_{n \times n},-\m I_{n \times n}\right\}.
\end{align}
The sets $\mathcal{T}^c$ and $\mathcal{T}^o$ represent the decision to apply both positive and negative unit perturbations to each input (for controllability covariance) and state (for observability covariance), respectively. This choice ensures that the system's response to both increases and decreases in inputs or initial states is evaluated, providing a comprehensive view of its dynamic behavior. Similarly, the sets $\mathcal{M}^c$ and $\mathcal{M}^o$ are chosen to be:
\begin{align}\label{eq:M set}
	\mathcal{M}^c=\alpha_u \mathcal{M}_0, \;\;\;\;\;
	\mathcal{M}^o =\alpha_x \mathcal{M}_0
\end{align}
where $\mathcal{M}_0=\{0.25,0.5,0.75,1.0\}$ is a linearly scaled set and offers a structured approach to varying the magnitude of perturbations. The set $\mathcal{M}_0$ ensures that the system's response is observed under perturbations ranging from subtle to substantial \cite{lall1999empirical, lall2002subspace}. In \eqref{eq:M set} an extra user-defined scaling constants $\alpha_u$ and $\alpha_x$ are also included to make the perturbation magnitude in the control input and states reasonable and to make sure the \texttt{ode15s} solver is able to simulate the system under those conditions. Notice that if the perturbation magnitude is too large, then the power system will lose synchronism, and the time-domain simulation will not be performed for those transient conditions. Thus, $\alpha_u$ and $\alpha_x$ need to be adjusted to make sure the system runs smoothly with \texttt{ode15s} solver. Given that, here we select the value for both of these constants to be $\alpha_u = \alpha_x = 0.05$, which works well in our case.
\vspace{-0.01cm}
\subsection{Case Study on IEEE 39-bus Systems}
\vspace{-0.01cm}
\begin{figure}[]
	\hspace{-0.0cm}\subfloat{\includegraphics[keepaspectratio=true,scale=0.51]{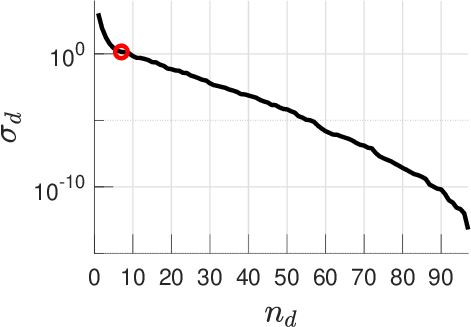}}{}{}\hspace{0.2cm}\subfloat{\includegraphics[keepaspectratio=true,scale=0.51]{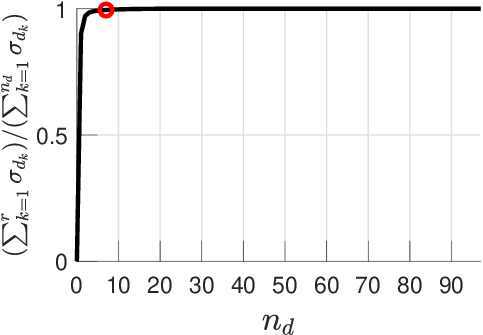}}{}{}\vspace{-0.1cm}
	
	\hspace{-0.0cm}\subfloat{\includegraphics[keepaspectratio=true,scale=0.51]{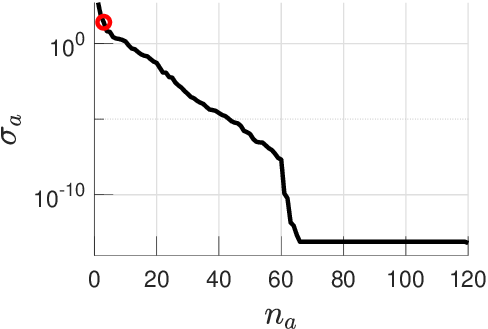}}{}{}\hspace{0.2cm}\subfloat{\includegraphics[keepaspectratio=true,scale=0.51]{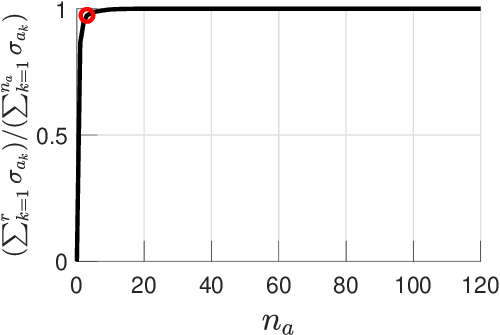}}{}{}\vspace{-0.3cm}
	\caption{HSVs and their cumulative sum contained in $\m\Sigma_d$ (above) while below is for $\m\Sigma_a$; 39-bus system. The first $r_d= 7$ HSVs in $\m\Sigma_d$ contain $99\%$ of the cumulative sum, similarly for $\m\Sigma_a$ the first $r_a= 3$ HSVs contain $97\%$ of the cumulative sum. Thus, the size of ROM is selected to be $r = 10$. }\label{fig:HSVs POD 39-bus}\vspace{-0.6cm}
\end{figure}
\begin{figure}[]
	\hspace{-0.0cm}\subfloat{\includegraphics[keepaspectratio=true,scale=0.51]{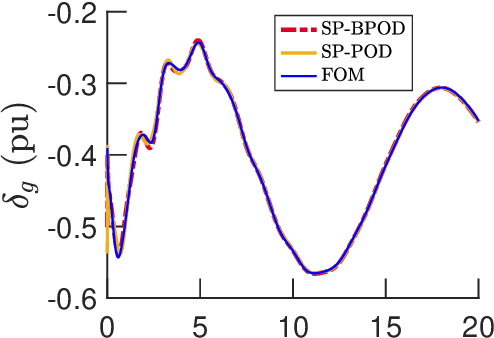}}{}{}\hspace{0.2cm}\subfloat{\includegraphics[keepaspectratio=true,scale=0.51]{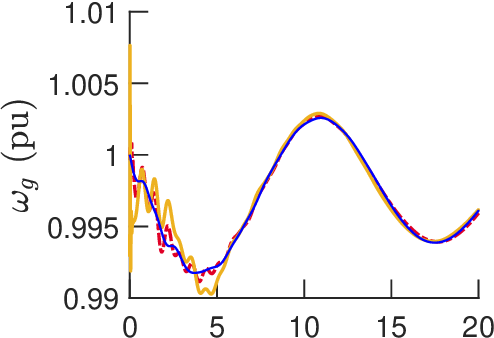}}{}{}
	
	\hspace{-0.0cm}\subfloat{\includegraphics[keepaspectratio=true,scale=0.51]{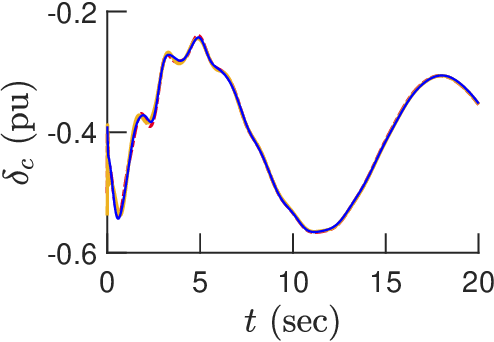}}{}{}\hspace{0.2cm}\subfloat{\includegraphics[keepaspectratio=true,scale=0.51]{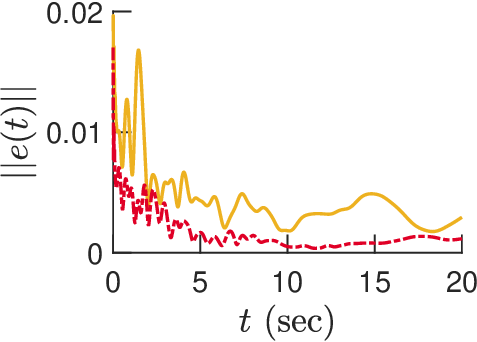}}{}{}\vspace{-0.3cm}
	\caption{Comparison of FOM and ROM for 39-bus system; rotor angle of Gen. at Bus $35$  (top-left), frequency of Gen. at Bus 35 (top-right), relative angle of solar plant  (bottom left), and overall error norm (bottom right).}\label{fig:plots 39 states}\vspace{-0.6cm}
\end{figure}
	\begin{figure}[htp!]
	\hspace{-0.0cm}\subfloat{\includegraphics[keepaspectratio=true,scale=0.51]{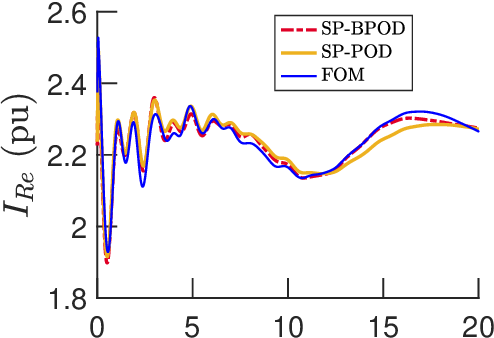}}{}{}\hspace{0.2cm}\subfloat{\includegraphics[keepaspectratio=true,scale=0.51]{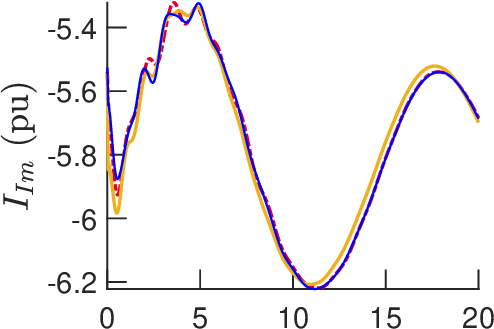}}{}{}
	
	\hspace{-0.0cm}\subfloat{\includegraphics[keepaspectratio=true,scale=0.51]{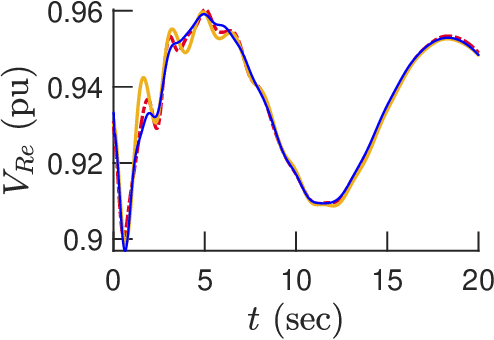}}{}{}\hspace{0.2cm}\subfloat{\includegraphics[keepaspectratio=true,scale=0.51]{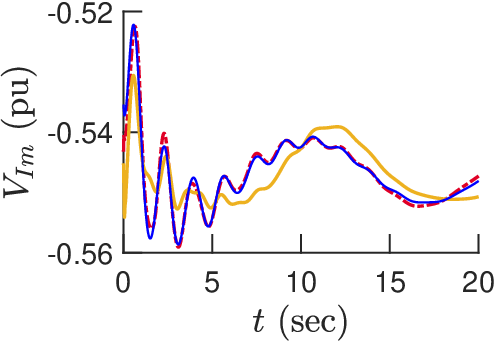}}{}{}\vspace{-0.3cm}
	\caption{Comparison of algebraic variables between FOM and ROM for the 39-bus system; Bus 5 real and imaginary current (above), and Bus 5 real and imaginary voltage (below).}\label{fig:plots 39 algebric states}\vspace{-0.2cm}
\end{figure}

Here, we perform MOR on a modified IEEE 39-bus systems. The 39-bus system consists of 9 conventional power plants (steam and hydro-based) and one solar farm. The conventional power plants are modeled via detailed $9^{th}$-order dynamics modeling; generator swing equations, turbine and governor models, and excitation system dynamics. The solar power plant is acting in grid-forming mode and is modeled via $12^{th}$-order dynamical model. Further details about the dynamics of the considered power system model are given in Appendix \ref{appndix:ninth Gen_dynamics}. The original full 39-bus system consists of $97$ dynamic states and $120$ algebraic variables, i.e $n_d = 97$, $n_a = 120$, and $n = 217$.

To design appropriate ROM for the considered test system, first we observe the HSVs contained in $\m\Sigma_d$, $\m\Sigma_a$ for SP-POD and $\m \Gamma_1$, $\m\Sigma_{gc}$ for SP-BPOD. The results for  $\m\Sigma_d$ and $\m\Sigma_a$ are shown in Fig. \ref{fig:HSVs POD 39-bus}. From these figures we can see that the HSVs of $\m\Sigma_d$ decay quickly to zero. Thus, $99\%$ system energy for the dynamic states in the transformed coordinates can be captured by choosing only the first $r_d = 7$ states and truncating the rest of them. Hence the dimension of identity matrix $\mI_{dr}$ is chosen to be $ \mI_{dr}  \in \mathbb{R}^{7 \times 7}$.  Similarly, from Fig. \ref{fig:HSVs POD 39-bus} by looking at the HSVs contained in $\m\Sigma_a$ we can see that the number of algebraic variables in the transformed coordinates can be truncated to 3 (as again $97\%$ of system input-output energy can be captured using first three states) by setting $\mI_{ar} \in \mathbb{R}^{3 \times 3}$. This truncation of both dynamic and algebraic variables in the transformed coordinates effectively reduced the dimension of the system by retaining much of the system input-output behavior while truncating those with little to no contribution to system dynamics.  

Similar observations have been carried out for the SP-BPOD-based MOR technique and the dimensions of ROM have been determined to be $r_d = 8$, $r_a = 3$, and $r=11$. After determining suitable dimensions for reduced models, the corresponding ROMs are constructed using the proposed SP-POD and SP-BPOD-based techniques as given in Algorithms \ref{alg:Algorithm 1} and \ref{alg:Algorithm 2}. A comparison with BT-based MOR has also been presented in Tabs. \ref{tab:Table 1} and \ref{tab:Table 2}, as proposed in  \cite{freitas2008gramian,BT_LODE} to showcase the effectiveness of the proposed methodologies. Notice that BT-MOR is applied to LODE systems as it cannot handle NDAE dynamics. 

To access the performance of the proposed MOR techniques time-domain simulations are carried out under transient conditions. The dynamic response of the power system is generated by adding step disturbance in load demand right at the start of the simulation. The disturbance in load demand has been added by choosing $\Delta_d = 0.005$ as discussed in the previous section. Both the full-order model (FOM) and ROMs are simulated under these transient conditions and system responses are recorded. To compare the performance between FOM and ROMs the data generated from ROMs are transformed backed and original state vectors are recovered. The results are given in Fig. \ref{fig:plots 39 states} and \ref{fig:plots 39 algebric states}. For brevity, a couple of dynamic and algebraic variables are shown, we can see the recovered dynamic states (rotor angle and speed of synchronous generator at Bus $35$ and relative angle of solar plant) are close to the original system responses. Similarly, for the algebraic variables shown in Fig. \ref{fig:plots 39 algebric states} we can see that the original and recovered states are close and accurate to each other. To further evaluate the performance of the proposed MOR techniques, we also compute the root mean square error (RMSE)  between the original and the recovered states. Accordingly, the RMSE value for SP-POD is determined to be $0.038$, while for the SP-BOPD-based ROM, it is $0.0019$. We can see that RMSE values are small and thus the ROMs can accurately approximate the full-order system dynamics. These results are also corroborated by Fig. \ref{fig:plots 39 states}  where the error norm is plotted, we can see that the error is close to zero.

Moreover, to assess the simulation accuracy between the FOM and ROMs separately for state variables of conventional power plants, solar plant, and algebraic states, we define the following index:
\begin{align}
	\varepsilon_s=\sqrt{\frac{\sum_{j=1}^N \sum_{t=1}^{t_f}\left(x_{j, t}^{\mathrm{r}}-x_{j, t}^{\mathrm{FOM}}\right)^2}{Nt_f}}
\end{align}
where $t_f$ is the overall time period and $N$ is the number of state variables whose accuracy needs to be determined. These results are shown in Tab. \ref{tab:Table 1}, we can see that the proposed methods can accurately approximate both the dynamic and algebraic variables of the full-order system dynamics. 
\begin{figure}[]
	\centering
	\vspace{-0.2cm}
	\hspace{-0.0cm}\subfloat{\includegraphics[keepaspectratio=true,scale=0.51]{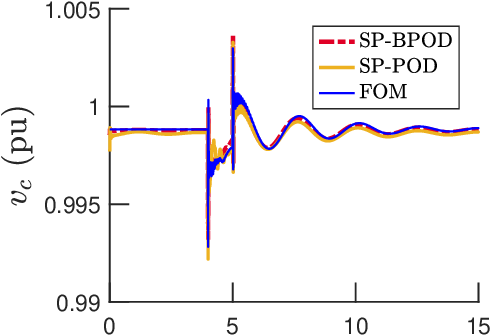}}{}{}\hspace{0.2cm}\subfloat{\includegraphics[keepaspectratio=true,scale=0.51]{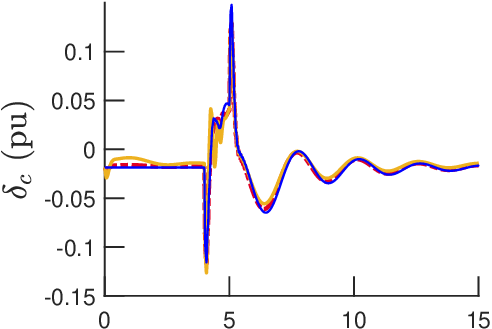}}{}{}
	
	\hspace{-0.0cm}\subfloat{\includegraphics[keepaspectratio=true,scale=0.51]{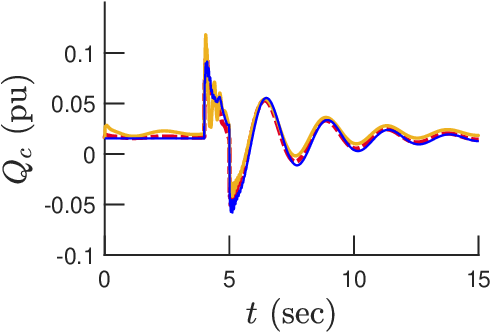}}{}{}\hspace{0.2cm}\subfloat{\includegraphics[keepaspectratio=true,scale=0.51]{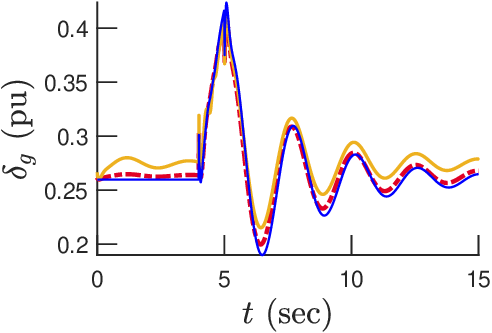}}{}{}
	\caption{Comparison of FOM and ROM for 39-bus system under fault; solar power plant voltage (top-left),  relative angle (top-right), reactive power output (bottom left), and rotor angle of Gen. at Bus 30 (bottom right).}\label{fig:mor fault}\vspace{-0.6cm}
\end{figure}

Here, we also asses the performance of the proposed technique under fault. To that end, we added a line to ground fault at $t = 4$sec on transmission line $4$-$14$ which is then cleared at $50$ msec and $200$ msec from the near and remote end. The simulation results are illustrated in  Fig. \ref{fig:mor fault} and we can see that both proposed techniques still yield accurate outcomes.
\vspace{-0.04cm}
\subsection{Case Study on 2000-bus Texas System}
\vspace{-0.01cm}
In this section, we discuss the performance of the proposed MOR techniques on a much larger power system model, namely, the 2000-bus Texas system.  The static network data (topology and parameters) on this system is taken from  \cite{Texas_system} while the dynamic data (the generator parameters) is generated synthetically. The overall system consists of $282$ synchronous machines modeled via a detailed $11^{th}$-order dynamical model. The dynamics consist of $6^{th}$-order generator swing equations, excitation system models, and turbine/governor dynamics. The original full-order system consists of $3102$ dynamic variables and $564$ algebraic variables, i.e., $n_d = 3102$, $n_a = 564$, and $n= 3666$.  

\begin{figure}[]
		\hspace{-0.0cm}\subfloat{\includegraphics[keepaspectratio=true,scale=0.51]{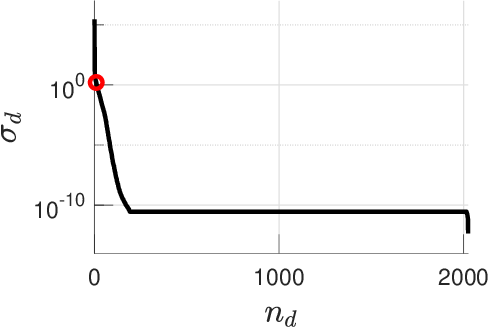}}{}{}\hspace{0.2cm}\subfloat{\includegraphics[keepaspectratio=true,scale=0.51]{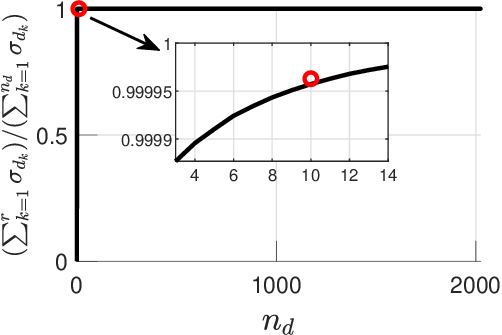}}{}{}\vspace{-0.1cm}
		
		\hspace{-0.0cm}\subfloat{\includegraphics[keepaspectratio=true,scale=0.51]{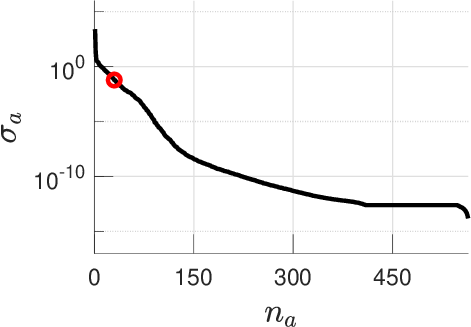}}{}{}\hspace{0.2cm}\subfloat{\includegraphics[keepaspectratio=true,scale=0.51]{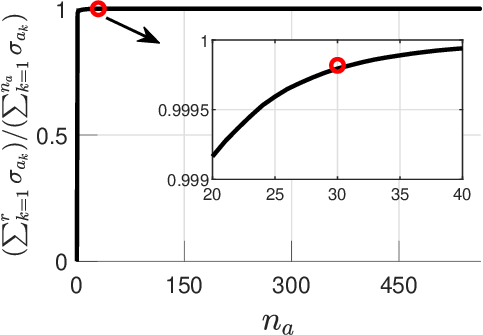}}{}{}\vspace{-0.3cm}
		\caption{HSVs and their cumulative sum contained in $\m\Sigma_d$ (above) while below is for $\m\Sigma_a$, 2000-bus Texas system. The first $r_d= 10$ HSVs in $\m\Sigma_d$ contain $99.99\%$ of the cumulative sum, similarly for $\m\Sigma_a$ the first $r_a= 30$ HSVs contain $99.99\%$ of the cumulative sum. Thus, the size of ROM is selected to be $r = 40$.}\label{fig:HSVs POD Texas-bus}\vspace{-0.4cm}
	\end{figure}

Now, similarly to as done in the previous section, to design ROM, we first determine the appropriate size of the dynamic and algebraic variables in the reduced order model NDAE. For the SP-POD this can be done by observing the HSVs in $\m\Sigma_d$ and $\m\Sigma_a$ as presented in Fig. \ref{fig:HSVs POD Texas-bus}. We see that the dynamic variables in ROM for the SP-POD-based technique can be set to be $r_d = 10$ (as this captures almost $99.99\%$ cumulative sum), similarly the number of algebraic variables in the ROM can chosen to be $r_a = 30$. This gives us the overall dimension for the SP-POD-based ROM as $r=40$. Similarly, for the SP-BPOD by observing HSVs in $\m \Gamma_1$ and $\m\Sigma_{gc}$ we get $r_d = 8$, $r_a = 25$, and $r=33$.

We want to point out here that while applying SP-BPOD for this case study, the computation of observability covariance $\m G_{0_{11}}$  can become computationally expensive. This is because to compute $\m G_{0_{11}}$ as discussed earlier, we have to perturb each dynamic state independently, and thus, we need to perform $n_d $ number of simulation studies. Which can become computationally expensive for the given Texas system with $n_d = 3102$.  However, this can be avoided, as discussed in \cite{rowley2005model}. The idea is only to perturb the dominant POD modes instead of all the dynamic states to approximate $\m G_{0_{11}}$. Additionally, one can consider bypassing the observability covariance computation altogether, relying on controllability alone to derive the ROM as in \cite{hahn2003controllability}.
 
\begin{figure}[htp!]
		\hspace{-0.0cm}\subfloat{\includegraphics[keepaspectratio=true,scale=0.51]{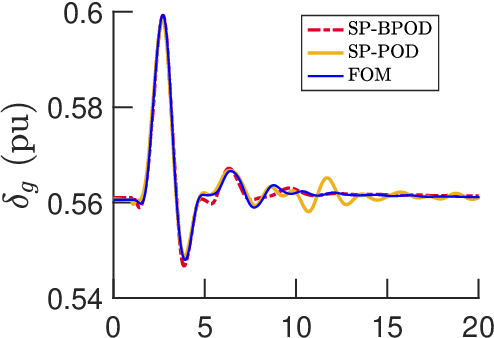}}{}{}\hspace{0.2cm}\subfloat{\includegraphics[keepaspectratio=true,scale=0.51]{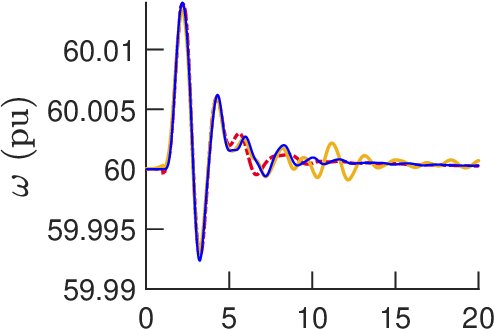}}{}{}
		
		\hspace{-0.0cm}\subfloat{\includegraphics[keepaspectratio=true,scale=0.51]{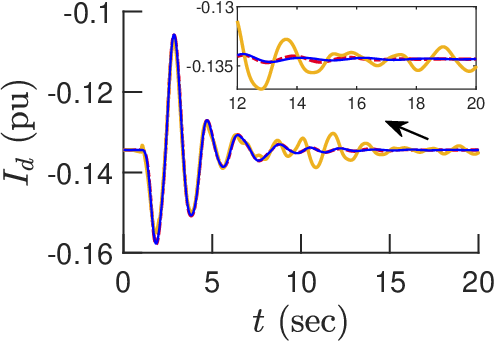}}{}{}\hspace{0.2cm}\subfloat{\includegraphics[keepaspectratio=true,scale=0.51]{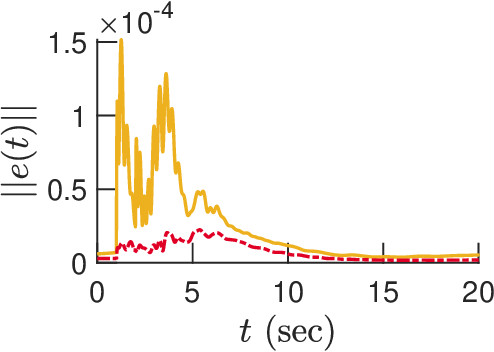}}{}{}\vspace{-0.4cm}
		\caption{Comparison of FOM and ROM for 2000-bus Texas system; Generator 10 rotor angle (top-left), frequency (top-right), current output (bottom left), and overall error norm (bottom right).}\label{fig:plots texas states}\vspace{-0.5cm}
	\end{figure}

Having said that, to assess the performance of the proposed technique, we again do time-domain simulations under transient conditions. To generate system dynamic response here, we create a generator side disturbance by adding a $10\%$ reduction in the mechanical power output of one of the synchronous generators at $t=1$s that last for $1$s. Both the FOM and ROMs are simulated under this transient condition and system responses are recorded. These results are shown in Fig. \ref{fig:plots texas states}. We can see that the response of both dynamic and algebraic variables from the ROMs closely matches that of the FOM. This can also be corroborated from the plot of error norm (presented in Fig.  \ref{fig:plots texas states} (bottom right)) and from the RMSE value which is determined to be $0.0037$ for SP-POD and $0.0027$. We can see that the error norm and the RMSE values are small, showing that the response from ROMs is accurate.

We also present the simulation accuracy for the dynamic and algebraic variables separately for this case study in Tab. \ref{tab:Table 2}. Again, we can verify that the proposed MOR techniques can simultaneously reduce both dynamic and algebraic variables (from $n_d = 3102$, $n_a = 564$, and $n= 3666$ to $r_d = 10$, $r_a = 30$, and $r= 40$ in case of SP-POD and to $r_d = 11$, $r_a = 31$, and $r= 42$ for SP-BPOD) while maintaining very good accuracy.  A comparison with state-of-the-art BT-based MOR has also been provided in Tab. \ref{tab:Table 2}. Notice that to apply BT the system has first been linearized and then converted to an equivalent ODE as in \cite{freitas2008gramian}. From Tab. \ref{tab:Table 2} we can see that the proposed MOR techniques are superior in terms of accuracy and the size of corresponding ROM as compared to BT-LODE. Notice that the reason BT-LODE is providing poor results is because it is designed based on linearized system dynamics, while the proposed techniques consider the complete nonlinear system. Also, BT-LODE cannot reduce the algebraic variables as it is applicable to ODE dynamics only, while the proposed MOR techniques can simultaneously reduce both dynamic and algebraic variables as presented in Tab. \ref{tab:Table 2}.

Finally, here we would like to compare and summarize the pros and cons of the two proposed MOR techniques.  Notice that, both the proposed SP-POD and SP-BPOD are general input-output-based MOR techniques that can be applied to any NDAE power system model represented in state-space format. Both techniques provide accurate ROMs but have their own advantages and limitations.
	
SP-POD is effective in capturing the dominant behaviors of a system by focusing on the energy content of the modes. However, it does not directly consider the impact of these modes on the system’s controllability or observability. This can be a critical limitation in control applications, as modes with lower energy levels might still significantly influence the system's response to controls and its observability. On the other hand, SP-BPOD accounts for the controllability and observability of the modes. However, computing the empirical controllability and observability Gramians (as detailed in Eqs. \eqref{eq:Wc} and \eqref{eq:Wo} can be very challenging, requiring systematic perturbation of control inputs (for the controllability Gramian) and states (for the observability Gramian) with fine-tuning, as explained in Sec. \ref{sec:case studies}. That being said, if the overall aim is to perform realtime scalable feedback control and monitoring, SP-BPOD may be the better option. Otherwise, SP-POD could be chosen due to its lesser computational complexity.
\begin{table}[]
	\setlength{\tabcolsep}{1pt}
	\centering
	\caption{Comparison of simulation accuracy of the proposed methods for various state variables, 39-bus system. The notation $r_d$ represents dynamic states in the ROM, similarly $r_a$ are the algebraic variables in the ROM, while $r$ denotes the overall dimension of ROM NDAE.}\vspace{-0.3cm}
	\label{tab:Table 1}
	\begin{tabular}{|c|ccc|}
		\hline
		\multirow{2}{*}{Variables}                                   & \multicolumn{3}{c|}{$\varepsilon_s=\sqrt{\frac{\sum_{j=1}^N \sum_{t=1}^{t_f}\left(x_{j, t}^{\mathrm{r}}-x_{j, t}^{\mathrm{FOM}}\right)^2}{N t_f}}$}                                                                                                                                                                                                                       \\ \cline{2-4} 
		& \multicolumn{1}{c|}{\begin{tabular}[c]{@{}c@{}}SP-POD\\$r_d=7$, $ r_a=3$\\$ r=10$ \end{tabular}} & \multicolumn{1}{c|}{\begin{tabular}[c]{@{}c@{}}SP-BPOD\\ $r_d=8$, $ r_a=3$,\\$r=11$\end{tabular}} & \begin{tabular}[c]{@{}c@{}}BT-LODE \cite{freitas2008gramian}\\  $r_d=12$\end{tabular} \\ \hline
		\begin{tabular}[c]{@{}c@{}}Conventional\\ Power plant States\end{tabular}   & \multicolumn{1}{c|}{$1.62\times 10^{-2}$}                                                                                         & \multicolumn{1}{c|}{$2.15\times 10^{-3}$}                                                                                          & $1.8813$                                                                                     \\ \hline
		\begin{tabular}[c]{@{}c@{}}Solar Farm\\ States\end{tabular} & \multicolumn{1}{c|}{$1.01\times 10^{-3}$}                                                                                         & \multicolumn{1}{c|}{$3.12\times 10^{-4}$}                                                                                          & $1.9412$                                                                                     \\ \hline
		\begin{tabular}[c]{@{}c@{}}Algebric \\ States\end{tabular}   & \multicolumn{1}{c|}{$4.01\times 10^{-3}$}                                                                                         & \multicolumn{1}{c|}{$1.58\times 10^{-4}$}                                                                                          & \begin{tabular}[c]{@{}c@{}}Not\\ Applicable\end{tabular}                                     \\ \hline
	\end{tabular}
\vspace{-0.2cm}
\end{table}
\begin{table}[]
	\setlength{\tabcolsep}{1pt}
	\centering
	\caption{Comparison of simulation accuracy of the proposed methods for various state variables, 2000-bus Texas system.}\vspace{-0.3cm}
	\label{tab:Table 2}
	\begin{tabular}{|c|ccc|}
		\hline
		\multirow{2}{*}{Variables}                                                & \multicolumn{3}{c|}{$\varepsilon_s=\sqrt{\frac{\sum_{j=1}^N \sum_{t=1}^{t_f}\left(x_{j, t}^{\mathrm{r}}-x_{j, t}^{\mathrm{FOM}}\right)^2}{Nt_f}}$}                                                                                                                 \\ \cline{2-4} 
		& \multicolumn{1}{c|}{\begin{tabular}[c]{@{}c@{}}SP-POD\\ $r_d\hspace{-0.0cm}=\hspace{-0.0cm}10$, $ r_a\hspace{-0.0cm}=\hspace{-0.0cm}30$\\ $r=40$\end{tabular}} & \multicolumn{1}{c|}{\begin{tabular}[c]{@{}c@{}}SP-BPOD\\ $r_d\hspace{-0.0cm}=\hspace{-0.0cm}11$, $ r_a\hspace{-0.0cm}=\hspace{-0.0cm}31$\\ $r=42$\end{tabular}} & \begin{tabular}[c]{@{}c@{}}BT-LODE \cite{freitas2008gramian}\\ $r_d\hspace{-0.0cm}=\hspace{-0.0cm}54$\end{tabular} \\ \hline
		\begin{tabular}[c]{@{}c@{}}Conventional\\Power Plant States\end{tabular} & \multicolumn{1}{c|}{$2.13\times 10^{-4}$}                                                        & \multicolumn{1}{c|}{$5.91\times 10^{-5}$}                                                         & $4.0828$                                                    \\ \hline
		\begin{tabular}[c]{@{}c@{}}Algebric \\ States\end{tabular}                & \multicolumn{1}{c|}{$7.91\times 10^{-4}$}                                                        & \multicolumn{1}{c|}{$0.87\times 10^{-5}$}                                                         & \begin{tabular}[c]{@{}c@{}}Not\\ Applicable\end{tabular}    \\ \hline
	\end{tabular}
\vspace{-0.4cm}
\end{table}

% 	\begin{figure}[htp!]
% 	\hspace{-0.0cm}\subfloat{\includegraphics[keepaspectratio=true,scale=0.51]{figures/Plots_Efd_TPWRS_tex.eps}}{}{}\hspace{0.2cm}\subfloat{\includegraphics[keepaspectratio=true,scale=0.51]{figures/Plots_Si_TPWRS_tex.eps}}{}{}
	
% 	\hspace{-0.0cm}\subfloat{\includegraphics[keepaspectratio=true,scale=0.51]{figures/Plots_Vr_TPWRS_tex.eps}}{}{}\hspace{0.2cm}\subfloat{\includegraphics[keepaspectratio=true,scale=0.51]{figures/Plots_Tm_TPWRS_tex.eps}}{}{}\vspace{-0.4cm}
% 	\caption{Comparison of conventional power plant states between FOM and ROM for 2000-bus Texas system; field voltage (top-left), damper winding flux linkage (top-right), exciter input (bottom left), and mechanical torque (bottom right).}\label{fig:plots texas Gen 10 states}\vspace{-0.6cm}
% \end{figure}
\vspace{-0.03cm}
\section{Concluding Remarks}\label{sec: conclusion}
\vspace{-0.01cm}
In this paper, we propose two MOR approaches, namely the SP-POD and SP-BPOD, to simultaneously reduce both dynamic and algebraic variables of NDAE power system models. The SP-POD offers reducing the system order based on the POD modes in the transient simulation data, while in contrast, the SP-BPOD offers designing ROM via balanced-realization. Because of the diagonal structure of the designed coordinate transformation matrix, the corresponding ROMs from both the proposed methods are guaranteed to be NDAE similar to the original power system model. Thus, the proposed techniques preserve the essential differential-algebraic structure of power system models while allowing a smooth transition from reduced order to the full order dynamic and algebraic state variables. Regarding the limitations of the proposed work we want to mention here that both the presented MOR techniques operate in a data-driven manner and require system transient data of all the state variables from time-domain simulations. Thus, to maintain a refined and accurate reduced model representation it needs to be updated with every power flow or optimal power flow cycle. Also, in this study, we only considered the dynamics of PV power plants while the models of wind and other renewable resources are neglected.  Furthermore, in the proposed SP-BPOD method, the sets $\mathcal{M}^c$ and $\mathcal{M}^0$, along with the scalars $\alpha_u$ and $\alpha_x$, can be considered hyper-parameters that need to be tuned according to the specific test system and operator expertise, as there is no systematic method provided for selecting their values.

To verify the accuracy of the proposed techniques, simulations on modified IEEE 39-bus and 2000-bus Texas systems are carried out. The results show that the proposed techniques can significantly reduce the size of NDAE power system models while providing state trajectories close to those directly computed from running the full power system model. Future work will focus on using the ROMs to design scalable robust state-feedback controllers and state estimation algorithms, making control algorithms more amenable to large-scale power systems.

\bibliographystyle{IEEEtran}
\vspace{-0.0cm}
\bibliography{mybibfile}

\appendices
\section{Details of the power system dynamics}\label{appndix:ninth Gen_dynamics}
\vspace{-0.0cm}
Borrowing from \cite{nadeem2023robust}, here we present comprehensive details about the dynamics of power system model used in this paper. In particular, we present the dynamics of conventional power plants, dynamics of grid-forming solar power plants, load dynamics, and the algebraic power flow equations. 
\vspace{-0.0cm}
\subsection{Dynamics of Conventional Power Plants}
\vspace{-0.0cm}
We model the conventional power plants via $9^{th}$-order dynamical model. The overall dynamics consist of synchronous machines swing equations, steam/hydro turbine and governor differential equations, and  IEEE-type DC1 excitation system model, presented as follows \cite{sauer2017power,nadeem2023robust}:
\begin{itemize}[leftmargin=*]
	\begin{subequations} \label{eq:SynGen9th}
		\item Swing equations:	
		\begin{align}
			\begin{split}
				\hspace{-0.25cm}\dot{\delta}_{\mr{g}_i} &= \omega_{\mr{g}_i} - \omega_{0}\\ 
				\begin{split}
					\dot{\omega}_{\mr{g}_i} &=\dfrac{1}{2H_i}( T_{\mr{M}_i}\hspace{-0.0cm}-\hspace{-0.0cm}T_{\mr{e}_i})\;\;\;\mr{with}\;\;\; T_{\mr{e}_i}\hspace{-0.0cm} =\hspace{-0.0cm} {E}_{\mr d_i}i_{\mr d_i} + {E}_{\mr q_i}i_{\mr q_i} \end{split}\\ 
				\hspace{-0.25cm}\dot{E}_{\mr q_i} &= -\frac{1}{t_{\mr{qo}_i}}(E_{\mr q_i}-(x'_{\mr{q}_i}-x_{\mr{q}_i})i_{\mr d_i})\\
				\dot{E}_{\mr d_i} &= -\frac{1}{t_{\mr{do}_i}}(E_{\mr d_i}+(x'_{\mr{d}_i}-x_{\mr{d}_i})i_{\mr q_i} - E_{\mr{fd}_i}). 			
			\end{split}
		\end{align}
		\item Turbine and governor dynamics:	
		\begin{align}\label{eq:gen_tur/gov-dyn}
			\begin{split}
				\hspace{-0.25cm}\dot{T}_{\mr{M}_i} &=
				\left\{
				\begin{array}{ll}
					-\frac{1}{t_{\mr{ch}i}} (T_{\mr{M}_i}-P_{v_i}) & \mr{if \;\;\; thermal}\\
					-\frac{2}{t_{wi}} (T_{\mr{M}_i}-P_{v_i}+ t_{\mr{ch}i}\dot P_{v_i})& \mr{if \;\;\; hydro}
				\end{array}
				\right.\\
				\dot{P}_{v_i} &= -\frac{1}{t_{vi}}(P_{v_i}-P^*_{v_i} + \dfrac{\omega_i - 1}{R_{di}}).
			\end{split}
		\end{align}
		\item Excitation system dynamics:	
		\begin{align}\label{eq:gen_excit_dyn}
			\begin{split}
				\begin{split}
					\hspace{-0.25cm}\dot{E}_{\mr{fd}_i} \hspace{-0.0cm}&=\hspace{-0.0cm} \frac{-1}{t_{\mr{fd}i}}(k_{ei}+S_{ei}E_{\mr{fd}_i} - v_{ai}) \;\;\;\mr{with}\;\;\; S_{ei}\hspace{-0.0cm} =\hspace{-0.0cm} a_i\mr{e}^{b_iE_{\mr{fd}i}}\end{split}\\
				\dot{r}_{f_i} &= -\frac{1}{t_{fi}}(r_{f_i} - \dfrac{k_{fi}}{t_{fi}}E_{\mr{fd}_i})\\
				\dot{v}_{ai} &= -\frac{1}{t_{ai}}(v_{ai} -k_{ai}v_{ei}).
			\end{split}
		\end{align} 
	\end{subequations}
\end{itemize}
In the above model, $i \in$ $\mathcal{G}$, $ \omega_{\mr{g}_i}$ represents generator speed, $\delta_{\mr{g}_i}$ denotes generator rotor angle, $\omega_0$ is synchronous speed of the generator, $E_{\mr d_i}$, $E_{\mr q_i}$ are the  generator transient voltages along dq-axis, $x'_{\mr q_i}, x'_{\mr d_i}$ represents synchronous generator transient reactance while $x_{\mr q_i},x_{\mr d_i}$  are the reactances along dq-axis, respectively, $t_{\mr{do}_i}$, $t_{\mr{qo}_i}$ denotes the dq-axis open circuit time constants, $i_{\mr q_i}$, $i_{\mr d_i}$ represents synchronous machine currents along dq-axis,   ${T}_{\mr{M}_i}$,  ${T}_{\mr{e}_i}$ are the mechanical torque and the electrical torque of generator, respectively, $P_{v_i}$ is hydro/steam  turbine valve position, $P^*_{v_i}$ is the operator set point for the turbine valve position, $v_{ai}$ denotes amplifier voltage, $E_{\mr{fd}_i}$ represents synchronous machine field voltage, $r_{f_i}$ denotes the stabilizer output, $R_{di}$ is the droop constant for the  governor $(\mr{Hz}/\mr{pu})$, $H_i$ represents inertia constant ($\mr{pu} \times \mr{sec}$) of the generator, and $k_{ai}$, $k_{ei}$, $k_{fi}$,  are the amplifier, exciter, and stabilizer constant  gains, respectively.

Furthermore, in model \eqref{eq:SynGen9th}, $t_{wi}$, $t_{\mr{ch}i}$, $t_{vi}$, $t_{fi}$, $ t_{ai}$, and $t_{\mr{fd}i}$  are the time constants for generator field voltage, hydro/steam turbine valve position, amplifier, and stabilizer, respectively while the notation $S_{ei}$ denotes saturation function of synchronous generator field voltage with scalar constants $a_i$, $b_i$ as given in \cite{sauer2017power}. Similarly, $v_{ei}$ in the exciter dynamics \eqref{eq:gen_excit_dyn} is the voltage control error given as: $v_{ei} = V^*_i -V_i+ r_f - \dfrac{k_{fi}}{t_f}E_{\mr{fd}_i}$ with $V_i$ representing the synchronous machine terminal voltage and $V^*_i$ denoting grid operator voltage set point.

The input and overall state vectors for the conventional power plant models can then be expressed as follows:
\begin{subequations}
	\begin{align}
		\m u_G &= \bmat{\m P^{*\top}_v & \m V^{*\top}}^\top \in \mbb{R}^{2G} \label{eq:inputSynGen} \\
		\m x_G\hspace{-0.05cm} &= \hspace{-0.05cm}\bmat{\m \omega_{\mr{g}}^\top \; \m \delta_{\mr{g}}^\top \; \m E_{\mr q}^\top \;  \m E_{\mr d}^\top\;\m E_{\mr{fd}}^\top\;\m T_\mr{M}^\top\;\m P_{v}^\top\;\m r_{f}^\top\; \m v_{a}^\top}^\top \hspace{-0.2cm}\in\hspace{-0.08cm} \mbb{R}^{9G} \label{eq:stateSyncGen}.
	\end{align}
\end{subequations}
\vspace{-0.3cm}
\subsection{Dynamics of Grid-Forming Solar Power Plants} 
\vspace{-0.01cm}
We model the solar plant dynamics via $12^{th}$-order dynamical model as given in \cite{SoumyaITPWRS2022, WasynczukITPE1996}. The overall model describes a solar power plant acting in  grid-forming (GFM) mode and the dynamics include;  DC side differential equation (dynamic equations describing PV array DC link models), AC side dynamics (DC/AC inverter and LCL filter differential equations), and current/voltage regulators dynamical models presented as follows with $i \in$ $\mathcal{R}$:
\begin{itemize}[leftmargin=*]
	\begin{subequations} \label{eq:PVdyn}
		\item DC side dynamics:
		\begin{align}\label{eq:DC_link_dyn}
			\dot{E}_{\mr{dc}_i} \hspace{-0.0cm}=\hspace{-0.0cm} \dfrac{1}{B_{C_i}}\left( P_{\mr{pv}_i}\hspace{-0.0cm} -\hspace{-0.0cm} P_{{\mr c}_i} \right).
		\end{align}
		\item AC side dynamics:
		\begin{align}\label{eq:AC_dyn}
			\begin{split}
				\hspace{-0cm}\dot i_{\mr{df}_i} &= \dfrac{\omega_b}{X_{f_i}}\left( -r_{f_i}i_{\mr{df}_i}\hspace{-0cm}+\hspace{-0.0cm} \omega_{c_i}X_{f_i}i_{\mr{qf}_i} \hspace{-0.0cm}+\hspace{-0.0cm} v_{\mr{df}_i}- v_{\mr{do}_i}\right) \\
				\hspace{-0cm}\dot i_{\mr{qf}_i} &= \dfrac{\omega_b}{X_{f_i}}\left( -r_{f_i}i_{\mr{qf}_i}\hspace{-0.0cm}+\hspace{-0.0cm} \omega_{c_i}X_{f_i}i_{\mr{df}_i} \hspace{-0.0cm}+\hspace{-0.0cm} v_{\mr{qf}_i}- v_{\mr{qo}_i}\right)\\
				\hspace{-0cm}\dot v_{\mr{dc}_i} &= \dfrac{\omega_b}{B_{c_i}}\left( \omega_{c_i}B_{c_i}v_{\mr{qc}_i} + i_{\mr{df}_i} - i_{\mr{dg}_i}\right) \\
				\hspace{-0cm}\dot v_{\mr{qc}_i} &= \dfrac{\omega_b}{B_{c_i}}\left( \omega_{c_i}B_{c_i}v_{\mr{dc}_i} + i_{\mr{qf}_i} - i_{\mr{qg}_i}\right) \\
				\hspace{-0cm}\dot\delta_{c_i} \hspace{-0.0cm}&=\hspace{-0.0cm} \omega_b(\omega_{c_i} \hspace{-0.0cm}- \hspace{-0.0cm}\omega_0) \;\;\; \mr{with}\;\;\; \omega_{c_i} \hspace{-0.0cm}=\hspace{-0.0cm} 1\hspace{-0.0cm}-\hspace{-0.0cm}k_{p_i}({\tilde{P}}_{e_i}\hspace{-0.0cm}-\hspace{-0.0cm}P^*_{e_i})\\
				\hspace{-0cm}\dot{\tilde{P}}_{e_i}\hspace{-0.0cm} &=\hspace{-0.0cm} \dfrac{1}{\tau_{s_i}}(-\tilde{P}_{e_i} \hspace{-0.0cm}+\hspace{-0.0cm} {P}_{e_i} )\\
				\hspace{-0cm}\dot{\tilde{Q}}_{e_i} \hspace{-0.0cm}&=\hspace{-0.0cm} \dfrac{1}{\tau_{s_i}}(-\tilde{Q}_{e_i}\hspace{-0.0cm} + \hspace{-0.0cm}{Q}_{e_i} ).
			\end{split}
		\end{align}
		\item Voltage regulator dynamics:
		\begin{align}\label{eq:vol_reg}
			\begin{split}
				\dot z_{\mr{do}_i}\hspace{-0.0cm} &=\hspace{-0.0cm} \dfrac{\kappa_{\mr{pv}_i}}{\tau_{v_i}}(v^*_{\mr{do}_i}\hspace{-0.0cm}-\hspace{-0.0cm}v_{\mr{do}_i})\;\;\; \mr{with}\;\;\; v^*_{\mr{do}_i}\hspace{-0.0cm}=\hspace{-0.0cm} V^*_i\hspace{-0.0cm}+\hspace{-0.0cm}k_{d_i}i_{\mr{qg}_i}\\
				\dot z_{\mr{qo}_i}\hspace{-0.0cm} &=\hspace{-0.0cm} \dfrac{\kappa_{\mr{pv}_i}}{\tau_{v_i}}(v^*_{\mr{qo}_i}\hspace{-0.0cm}-\hspace{-0.0cm}v_{\mr{qo}_i})\;\;\; \mr{with}\;\;\; v^*_{\mr{qo}_i}\hspace{-0.0cm}=0.
			\end{split}
		\end{align}	 
		\item Current regulator dynamics:
		\begin{align}\label{eq:Currt_reg_dyn}
			\begin{split}
				\dot z_{\mr{df}_i}\hspace{-0.0cm} &=\hspace{-0.0cm} \dfrac{\kappa_{\mr{p}_i}}{\tau_{i_i}}(i^*_{\mr{df}_i}\hspace{-0.0cm}-\hspace{-0.0cm}i_{\mr{df}_i})\\
				i^*_{\mr{df}_i}\hspace{-0.0cm}&=\hspace{-0.0cm} \hspace{-0.0cm}\kappa_{\mr{pv}_i}(v^*_{\mr{do}_i}-v_{\mr{do}_i}+ z_{\mr{do}_i}+i_{\mr{dg}_i}+i_{\mr{dc}_i})\\
				\dot z_{\mr{qf}_i}\hspace{-0.0cm} &=\hspace{-0.0cm} \dfrac{\kappa_{\mr{p}_i}}{\tau_{i_i}}(i^*_{\mr{qf}_i}\hspace{-0.0cm}-\hspace{-0.0cm}i_{\mr{qf}_i})\\
				i^*_{\mr{qf}_i}\hspace{-0.0cm}&=\hspace{-0.0cm} \hspace{-0.0cm}\kappa_{\mr{pv}_i}(v^*_{\mr{qo}_i}-v_{\mr{qo}_i}+ z_{\mr{qo}_i}+i_{\mr{qg}_i}+i_{\mr{qc}_i}).
			\end{split}
		\end{align}
	\end{subequations}
\end{itemize}

\noindent In Eq. \eqref{eq:DC_link_dyn}, $B_{C_i}$ denotes the capacitance of the DC link capacitor while $E_{\mr{dc}_i}$ represents the energy stored in it, $P_{\mr{pv}_i}$ denotes the DC power supplied by the PV array while $P_{{\mr c}_i}$ represents the power extracted by the solar inverter.  

Similarly, in AC side dynamics \eqref{eq:AC_dyn}, $X_{f_i}$, $r_{f_i}$ denotes the reactance and resistance of the AC side LCL filter, $i_{\mr{df}_i}$, $i_{\mr{qf}_i}$ represents the current flowing through the LCL filter along dq-axis,  $B_{c_i}$, $r_{c_i}$  denotes the capacitance and resistance of the LCL filter capacitor,  $i_{\mr{dc}_i}$, $i_{\mr{qc}_i}$, $v_{\mr{dc}_i}$, $v_{\mr{qc}_i}$ represents the current and voltages of the LCL filter capacitor along dq-axis, respectively, $i_{\mr{dg}_i}$, $i_{\mr{qg}_i}$ are the dq-axis current output to the main grid, $\omega_b$ represents the base speed while $\omega_{c_i}$ denotes the angular speed of the inverter, $P_{e_i}$, $Q_{e_i}$ are the real and reactive power output of the solar plant to the main grid, $\tilde{P}_{e_i}$,  $\tilde{Q}_{e_i}$ are the phasor representations of ${P}_{e_i}$,  ${Q}_{e_i}$ after passing through low pass filter (which are later used in the droop control of the GFM inverter), $\tau_{s_i}$ represents the time constant of the low pass filter, $k_{p_i}$ denotes the droop constant of the solar inverter, and finally $P^*_{e_i}$ represents the grid operator active power set-point command. 

Furthermore, in voltage and current regulator dynamics given in \eqref{eq:vol_reg} and \eqref{eq:Currt_reg_dyn}, $V^*_i$ represents the grid operator voltage set point command, $k_{q_i}$ denotes the voltage droop constant, and $\kappa_{\mr{p}_i}$, $\kappa_{\mr{pv}_i}$, $\tau_{i_i}$, $\tau_{v_i}$  are the constants gains and corresponding time constants of current and voltage regulators of the solar plants, respectively. Note that, current and voltage regulation in the presented GFM inverter is simply achieved by a proportional-integral (PI) type controller with  $z_{df}$, $z_{qf}$, $z_{qo}$, $z_{do}$ representing the states of integral compensators along dq-axis, respectively, as detailed in \cite{WasynczukITPE1996, SoumyaITPWRS2022}.

Therefore, the overall input and state vector for the solar power plant model used in this study can be expressed as:
	\begin{align*}
		\m u_R &= \bmat{\m P^{*\top} & \m V^{*\top}}^\top\in\mbb{R}^{2R}\\
		\m x_R\hspace{-0.0cm} &=\hspace{-0.0cm} \bmat{\m \delta_{\mathrm{c}}^\top\;\;\;\m E_{\mr{dc}}^\top\;\;\m P_{e}^\top\;\;\m Q_{e}^\top\;\;\m i_{\mr{dqf}}^\top\;\;\m v_{\mr{dqc}}^\top \;\;\m z_{\mr{dqo}}^\top\;\;\m z_{\mr{dqf}}^\top}^\top\hspace{-0.0cm} \in \hspace{-0.0cm}\mbb{R}^{12R}.
	\end{align*}
\vspace{-0.0cm}
\subsection{Power System Algebraic Equations and Loads Dynamics}
\vspace{-0.0cm}
Here, we present the algebraic constraints and the load models of the considered test power system. We consider various types of loads dynamics  $i \in$ $\mathcal{L}$ such as constant impedance, constant power, and motor type loads detailed as follows \cite{nadeem2023robust}.

The differential equations for the motor-based loads are given as:
\begin{equation}
	\dot{\omega}_{\mr{M}_i} = \frac{1}{2H_{\mr M_i}}(T_{{e_i}} - T_{\mr M_i}) \label{eq:omegam}
\end{equation}
where $H_{\mr M_i}$ denotes the inertia constant of the motor, $\omega_{\mr M_i}$ represents the speed of the motor-based load, and $T_{{e_i}}$, $T_{{M_i}}$ denotes the electromagnetic and mechanical torque of the motor, respectively. 

Constant impedance and constant power types loads satisfy the following relationships \cite{nadeem2023robust}: 
\begin{subequations}
	\begin{align}
		I_{z_i}Z_i+	V_{z_i} &= 0\\
		\begin{split}\label{eq:load_dyn_z}
		P_{p_i}+Q_{p_i} + \mr{conj}(I_{p_i})V_{p_i} &= 0
		\end{split}
	\end{align}
\end{subequations}
where $	\mr{conj}$ denotes complex conjugate operator, and $P_{p_i}$, $Q_{p_i}$, $V_{p_i}$, $I_{p_i}$ are the real power, reactive power, voltage, and current phasors of buses connected to constant power loads, respectively. Similarly, $I_{z_i}$, $V_{z_i}$ are the current and voltage phasors of the buses connected to the constant impedance loads $Z_i$.

The algebraic constraints are the current balance equations and are given as follows:
\begin{gather}
	\underbrace{\begin{bmatrix}
			\m{{I}}_{R} \\ \m{{I}}_{G} \\ \m{{I}}_{L}
	\end{bmatrix}}_{\m I(t)}
	-
	\underbrace{\begin{bmatrix}
			\m{Y}_{RG} & \m{Y}_{RR} & \m{Y}_{RL} \\
			\m{Y}_{GG} & \m{Y}_{GR} & \m{Y}_{GL} \\
			\m{Y}_{LG} & \m{Y}_{LR} & \m{Y}_{LL} 
	\end{bmatrix}}_{\m Y}
	\underbrace{\begin{bmatrix}
			\m{{V}}_{R} \\ \m{{V}}_{G} \\ \m{{V}}_{L} \\
	\end{bmatrix}}_{\m V(t)} =  \m{0} \label{eq:transalgebraic}
\end{gather}
where $\m I(t)$ denotes the net injected current, $\m V(t)$ represents bus voltages,  and $\m{Y}$ is the power network admittance matrix. Moreover, $\m{{I}}_{G}\hspace{-0.13cm}= \hspace{-0.13cm}\{I_{Re_i}\}_{i\in \mc{G}}\hspace{-0.05cm}+\hspace{-0.05cm}j \{I_{Im_i}\}_{i\in \mc{G}}\hspace{-0.05cm}$ , $\m{{V}}_{G} \hspace{-0.06cm}= \hspace{-0.06cm}\{V_{Re_i}\}_{i\in \mc{G}}\hspace{-0.05cm}+\hspace{-0.05cm} j\{V_{Im_i}\}_{i\in \mc{G}}\hspace{-0.05cm}$ represents current and voltage phasors at the terminal of buses connected with conventional power plants. Similarly,
$\m{{I}}_{R}$, $\m{{I}}_{L}$, and $\m{{V}}_{R}$, $\m{{V}}_{L}$  are the current and voltage phasors of solar plants and load buses, respectively. 

Having said that, the overall state vectors for loads and system algebraic constraints  can be written as follows:
\begin{subequations}\label{eq:alg_stat}
	\begin{align}
		\m x_L &= \bmat{\m \omega_{\mr m}}\in\mbb{R}^{L_k}\\
		\m x_a &= \bmat{\m I_{Re}^\top&\m I_{Im}^\top&\m V_{Re}^\top&\m V_{Im}^\top}^\top \in\mbb{R}^{4N}.
	\end{align}
\end{subequations}

%\section{Proposed MOR algorithms}\label{appndix:alg}
%\noindent The proposed SP-POD and SP-BPOD-based MOR algorithms are detailed in this section. Further explanations about creating the snapshot matrices ($\mX_d$, $\mX_a$, and $\mX_f$) and covariance matrices ($\mG_{c_{11}}$  and  $\mG_{o_{11}}$) are given in Sec. \ref{sec:case studies}.

% \begin{algorithm}\label{alg:Algorithm 2}
% 	\caption{\text{SP-BPOD for power system NDAE models}}
% 	\DontPrintSemicolon 
% 	Compute controllability covariance $\mG_c$\;	
% 	Decompose $\mG_c$ and extract covariances $\mG_{c_{11}}$ and $\mG_{c_{22}}$ \;	
% 	Compute observability covariance $\mG_{o_{11}}$\;	
% 	Design $\mT_1$, $\mT_2$, $\mT_3$, and $\mT_4$ using $\mG_{c_{11}}$ and $\mG_{o_{11}}$\;
% 	Balance dynamic states using coordinate transformation 	$\mW_d = \mT_1\mT_2\mT_3\mT_4$\;
% 	Perform SVD of $\mG_{c_{22}}$ as:
% 	$\m G_{c_{22}} = \m W_{gc}\m\Sigma_{gc}\m \Lambda_{gc}$\;
% 	Set $\mW_a = \m W_{gc}$ \;
% 	Construct $\mW = \mr{blkdiag}(\mW_d,\; \mW_a)$\;
% 	Design $\mT_d$, $\mT_a$ by observing HSVs in  $\m\Gamma_1$ and $\m\Sigma_{gc}$\;
% 	Construct the truncation matrix $\mT = \mr{blkdiag}(\mT_d,\; \mT_a)$\;
% 	Design $\mW_L = \mT\mW$ and  $\mW_R = \mW_L^{-1}$\;
% 	Construct matrix $\mX_f$\;	
% 	Follow 'Part $2$' of Algorithm \ref{alg:Algorithm 1} to perform greedy-based hyper-reduction for the nonlinearity and to compute $\tilde{\m f}(\cdot)$\;
% 	Design $\mE_r$, $\mA_r$, $\mB_{ur}$, and $\mB_{wr}$ as in Eq. \eqref{eq:ROM_initial}
% \end{algorithm}

\begin{IEEEbiography}[{\includegraphics[width=1in,height=1.25in,clip,keepaspectratio]{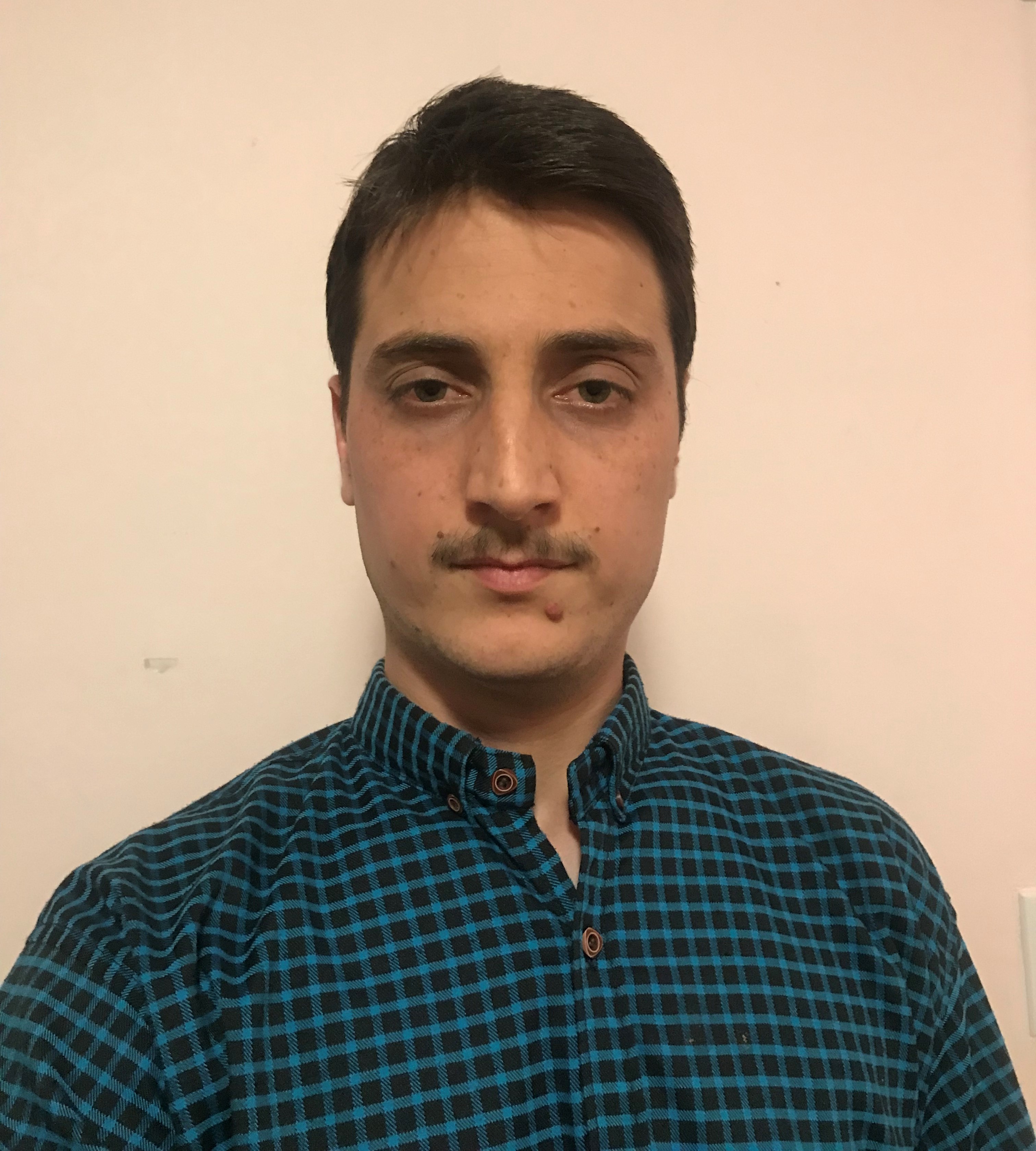}}]
	{Muhammad Nadeem} was born in Nilore, Islamabad, Pakistan. He received the B.E. and M.S. degrees in Electrical Engineering Power from Air University and the National University of Science and Technology (NUST) Islamabad Pakistan in 2017 and 2020. He is currently a graduate research assistant pursuing a Ph.D. degree in civil and environmental engineering at Vanderbilt University, Nashville, Tennessee, USA. His research interests include control theory, reinforcement learning, system identification, and state estimation for renewables integrated power systems.
\end{IEEEbiography}
\vspace{-0.6cm}
\begin{IEEEbiography}[
{
\includegraphics[width=1in,height=1.25in,clip,keepaspectratio]{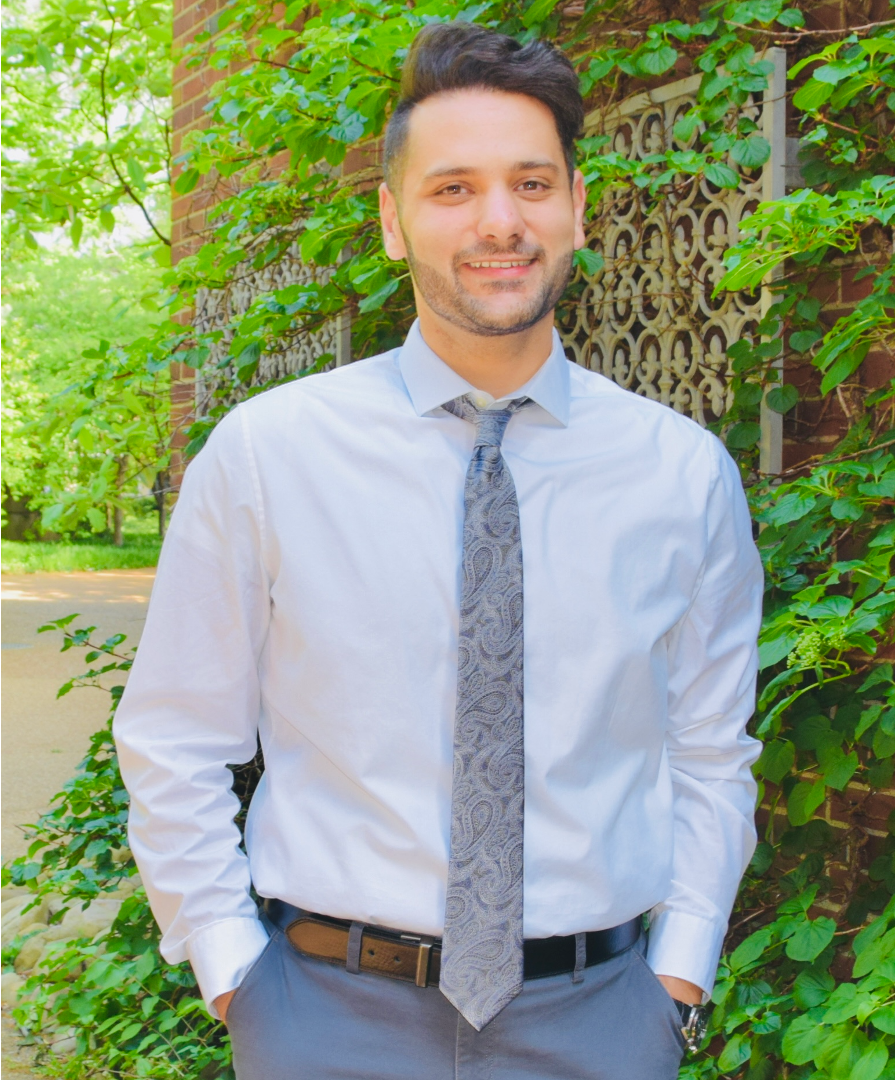}
}
]
{Ahmad F. Taha}  is an associate professor with the Department of Civil and Environmental Engineering at Vanderbilt University in Nashville, Tennessee. He has a secondary appointment in Electrical and Computer Engineering. He received his B.E. and Ph.D. degrees in Electrical and Computer Engineering from the American University of Beirut, Lebanon in 2011 and Purdue University, West Lafayette, Indiana in 2015. Prior to joining Vanderbilt. Taha was an assistant professor with the ECE department at the University of Texas, San Antonio (UTSA). Dr. Taha is interested in understanding how complex cyber‐physical, urban infrastructure operate, behave, and occasionally \textit{misbehave}. His research focus includes optimization, control, monitoring, and security of infrastructure with applications to power, water, and transportation systems. Dr. Taha is an associate editor of IEEE Transactions on Control of Network Systems. 
\end{IEEEbiography}

\end{document}